\documentclass[a4paper,fleqn]{cas-dc}

\usepackage[numbers]{natbib}

\def\tsc#1{\csdef{#1}{\textsc{\lowercase{#1}}\xspace}}
\tsc{WGM}
\tsc{QE}
\tsc{EP}
\tsc{PMS}
\tsc{BEC}
\tsc{DE}

\begin{document}
\let\WriteBookmarks\relax
\def\floatpagepagefraction{1}
\def\textpagefraction{.001}

\shorttitle{}

\shortauthors{Ashish Gupta et~al.}

\title [mode = title]{Calculation of true coincidence summing correction factor for a Broad Energy Germanium (BEGe) detector using standard and fabricated sources}                      
\tnotemark[]




%



\ead{anjali.mukherjee@saha.ac.in}


\credit{Conceptualization of this study, Methodology, Software}

\author[1,2]{Ashish Gupta}
\author[1]{M. Shareef}
\author[1,2]{Munmun Twisha}
\author[1,2]{Saikat Bhattacharjee}
\author[3]{Gopal Mukherjee}
\author[3]{Satya Samiran Nayak}
\author[3]{Sansaptak Basu}
\author[4]{S. Dasgupta}
\author[4]{J. Datta}
\author[3]{S. Bhattacharyya}
\author[1,2]{A. Mukherjee}[type=editor,
                        auid=000,bioid=1,
                        ]
                        
\cormark[1]
\affiliation[1]{organization={Saha Institute of Nuclear Physics},
            addressline={1/AF, Bidhan Nagar}, 
            city={Kolkata},
            postcode={700064}, 
            country={India}}

\affiliation[2]{organization={Homi Bhabha National Institute},
            addressline={Anushaktinagar}, 
            city={Mumbai},
            postcode={400094}, 
            country={India}}

\affiliation[3]{organization={Variable Energy Cyclotron Centre},
            addressline={1/AF, Bidhan Nagar}, 
            city={Kolkata},
            postcode={700064}, 
            country={India}}
    
\affiliation[4]{organization={Analytical Chemistry Division, Bhabha Atomic Research Centre, Variable Energy Cyclotron Centre},
            addressline={1/AF, Bidhan Nagar}, 
            city={Kolkata},
            postcode={700064}, 
            country={India}}
    
\cortext[cor1]{Corresponding author}



\begin{abstract}
The true coincidence summing (TCS) correction factor for a Broad Energy Germanium (BEGe) detector has been calculated at far and close geometry measurement using multi-energetic radioactive $\gamma$-ray sources $^{60}$Co, $^{133}$Ba and $^{152}$Eu. The correction factors were calculated using experimental method and analytical method. Photopeak efficiency and total efficiency  required to calculate the correction factor were obtained using Geant4 Monte Carlo simulation code. A few standard as well as fabricated mono-energetic sources were also included in the $\gamma$-ray efficiency measurements. The simulated efficiencies of mono-energetic $\gamma$-ray sources were matched to experimental $\gamma$-ray efficiencies by optimizing the detector parameters. The same parameters were used to obtain the photopeak and total efficiency for $\gamma$-ray of our interest and coincident $\gamma$-ray. Analytical correction factors and experimental correction factors were found in good agreement with each other.
\end{abstract}


\begin{highlights}
\item Close geometry measurement for $\gamma$-ray detection efficiency
\item Coincidence summing correction for BEGe (Broad Energy Germanium) detector

\end{highlights}


\maketitle

\section{Introduction}
\label{sec:Introduction}

In nuclear astrophysics arena, the nuclear reactions of stellar evolution occurs within the Gamow window which lies much below the coulomb barrier. The cross-sections of these reactions are very less $\sim nb-pb$ {\color{blue}\textbf{\cite{Bruno2018,Descouvemont2020}}}, thereby making the measurements very difficult and therefore requires extra care in performing the experiments. In such measurements, the resulting low yields of the $\gamma$-rays originating from the decay of the excited states to ground states of the product nuclei, lead to insufficient statistics and hence increase the uncertainty in the measurement of the $\gamma$-ray cross-sections. The $\gamma$-ray cross-sections are obtained from the detected yield of the $\gamma$-ray, using the relation:

\begin{equation}\label{cross-section}
    \sigma = \frac{Y}{\epsilon N_b N_t}
\end{equation}

where, $Y$, $\epsilon$, $N_b$ and $N_t$ are the yield, detector efficiency, beam current and number of target nuclei per unit area, respectively. A better statistics in the measurement may be achieved by  increasing the yield, which is proportional to $N_b$, $N_t$ and $\epsilon$. Among these factors, $N_b$ and $N_t$ can be increased, but there are practical limitations. It is also possible to enhance the efficiency of the detection system. One way to increase the efficiency is to place the source close to the face of the detector. However, such close geometry measurements may introduce coincidence summing effect in $\gamma$-ray efficiency measurements.

It is well known fact that many unstable nuclei de-excite to the ground states via the emission of multiple $\gamma$-rays from its higher energy states. When two or more $\gamma$-rays emitted in cascade from the excited nucleus are detected within the resolving time of the detector then this phenomenon is referred as the true coincidence summing (TCS).  This TCS is not a random summing, the latter effect is related to pulse pileup where photons of different cascades sum their energies randomly because of relatively high pulse rates. Where as TCS is independent of pulse rate but depends on the source to detector distance and the decay scheme {\color{blue}\textbf{\cite{IliadisBook}}} of the source. The origin of TCS observed is not only due to the simultaneous detection of two or more cascading $\gamma$-rays but may also be due to the simultaneous detection of a $\gamma$-ray and a X-ray {\color{blue}\textbf{\cite{HaquinPaper}}}. This X-ray may be emitted from internal conversion of any $\gamma$-ray from decay scheme or electron capture transitions.
    
Even though various modes for an excited nucleus are possible to decay to its ground state, three types of phenomenon may results in TCS. As an example, in the case of coincidence of two $\gamma$-rays, if the first $\gamma$-ray deposits all of its energy and the coincident $\gamma$-ray deposits only part of its energy in the detector, then there is a loss in the count from first $\gamma$-ray photopeak. This is referred as "summing-out". If both $\gamma$-rays deposit all of their energies in the detector, a sum peak appears in the spectrum, at an energy corresponding to the sum of the energies of the two $\gamma$-ray and this is referred as "summing-in". If a nucleus independently emits a $\gamma$-ray with an energy equal to the energy of the sum peak then we observe extra counts in the area under the curve of that photopeak energy. The simultaneous detection of two $\gamma$-rays which deposits full of its energy is less probable. Last case is when both the coincident $\gamma$-rays deposit only parts of their energies in the detector, it contributes only to the background and neither influence the peak intensity nor the efficiency.

\begin{figure}[t]
    \centering
    \includegraphics[scale = 0.55]{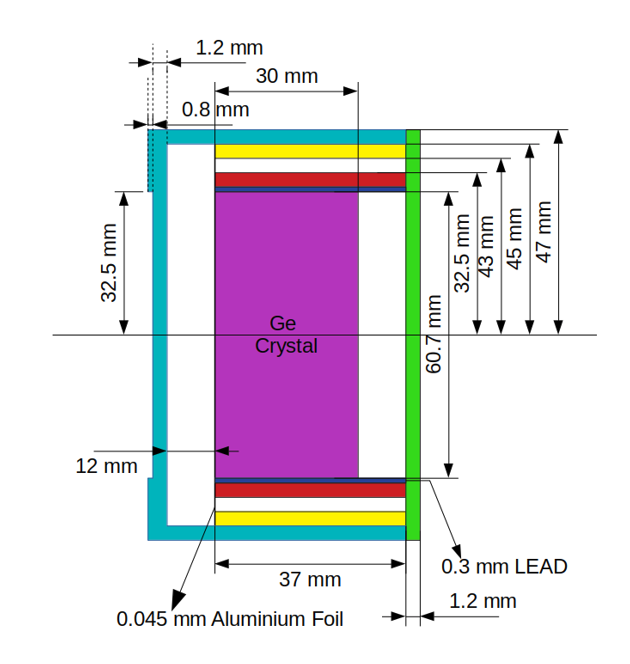}
    \caption{Schematic diagram of BEGe detector geometry with dimensions as provided by the manufacturer}
    \label{fig:BEGe detctor designe}
\end{figure}

In close geometry, the coincidence summing effect increases as the distance between the source and the detector decreases. This coincidence summing is an unwanted effect which contributes to the miscalculation of photopeak area and hence the photopeak efficiency. As cross-section in the astrophysical energy domain are very small, close geometry measurements become essential. Hence in such cases, coincidence summing correction for the detection system become necessary to have an accurate knowledge of detector efficiency. The method for calculating the coincidence summing correction factor was first described in {\color{blue}\textbf{Ref. \cite{Andreev1972}}} and later improved in {\color{blue}\textbf{Ref. \cite{Mccallum1975, Debertin1978, Semkov1990, Kourn2004}}}. The detailed procedure for calculating the correction factor with this method is described in {\color{blue}\textbf{Ref. \cite{DebrtinBook, Xilei1991}}}.

\begin{figure}
    \centering
    \includegraphics[scale = 0.35]{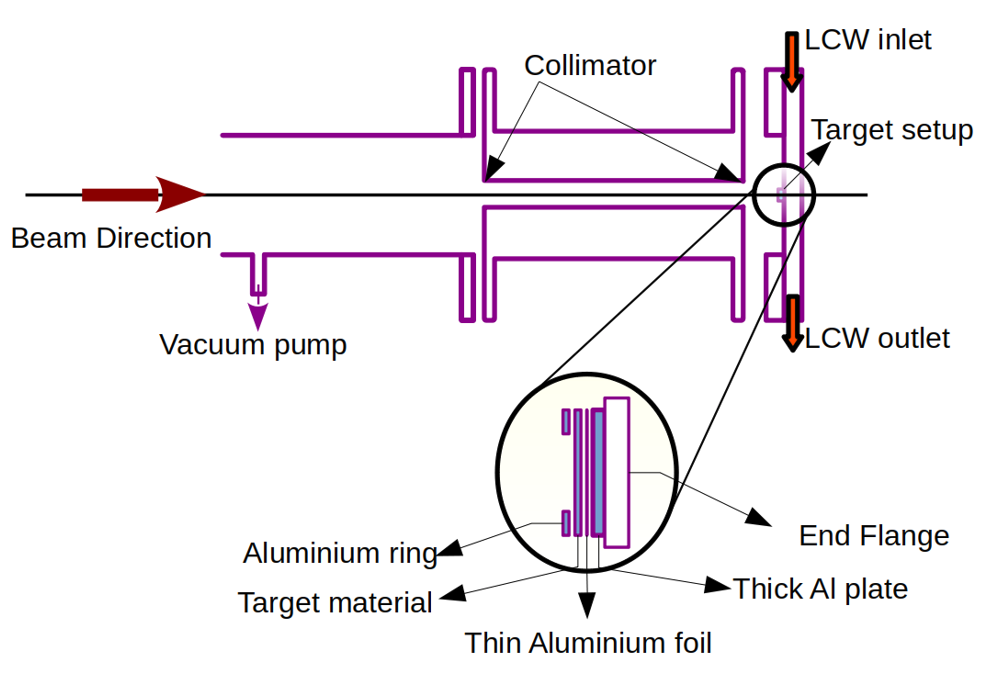}
    \caption{Experimental setup of activation method for mono-energetic source preparation at VECC, Kolkata}
    \label{fig:accelerator ecperimental setup}
\end{figure}

The method described here is an analytical approach to calculate the TCS correction factors. This method requires the photopeak efficiency, total efficiency and decay parameters, such as the mode of the parent nuclei decay, energies of the $\gamma$-transitions, $\gamma$-ray emission probability, K-capture probability (in $e^-$ capture decay), mean energies of the X-rays, fluorescence yield, total and K conversion coefficients. Information of all these factors are required to calculate the probability of simultaneous emission of two or more $\gamma$-rays. The correction factor can be calculated to desired accuracy and to a first order approximation the factor is given by {\color{blue}\textbf{\cite{ChhaviAgarwal2011}}},

\begin{equation}\label{correcton factor}
    k_{TCS} = \frac{1}{1-\Sigma_{i=1}^{i=n}p_i\epsilon_{ti}}
\end{equation}

where $n$ is number of $\gamma$-rays in coincidence with the $\gamma$-ray of interest, $p_i$ is the probability of simultaneous emission of $i^{th}~ \gamma$ and $\gamma$-ray of interest and $\epsilon_{ti}$ is the total efficiency of the $i^{th}~\gamma$-ray. The $\gamma$-ray photopeak efficiencies can be obtained over a wide energy region using multi-energetic $\gamma$-ray sources but the measurement of total efficiency is very complicated with multi-energetic radioactive sources since the spectrum cannot be decomposed into well defined components belonging to $\gamma$-rays with distinct energies. So we can only use the mono-energetic radioactive sources which emits single photopeak energy ($E_\gamma$). However, the availability of mono-energetic $\gamma$-ray sources are very limited and their half-lives ($t_{1/2}$) are often very small, around days, so we need to replace them periodically.

\begin{table}[]
    \centering
    \begin{tabular}{m{2.8cm} m{2.3cm} m{2.4cm} }
    \hline
    Detector parameters &   Manufacturer dimensions (mm) & Optimized dimensions (mm) \\
    \hline
     Crystal radius   &   30.35    &  29.45  \\
     Crystal length     &   30      & 29.5 \\
     Deal layer         &   0.045  & 1.8 \\
     Al end cap thickness   &   1.2 & 1.2  \\
     Al end cap to crystal distance ($d_{alc}$) & 12 & 14.5 \\
     \hline
    \end{tabular}
    \caption{Detector dimension parameters provided by the manufacturer and optimized by Geant4 Monte Carlo simulation}
    \label{tab:Detector parameters}
\end{table}

The other method is, to simulate the $\gamma$-ray spectrum at each energy of interest for a given detector geometry and calculate the photopeak and total efficiency from the simulated $\gamma$-ray spectrum. This simulation is independent of decay-scheme of radioactive source, hence its results are free from coincidence summing effect. The exact and precise knowledge of detector dimensions like crystal diameter, crystal length, dead layer thickness, aluminium end cap thickness and aluminium end cap to crystal distance are required to simulate the $\gamma$-ray spectrum. These parameters provided by the manufacturer may not be accurate or may change with the course of time. In a standard procedure, these detector parameters are optimized using measured efficiencies with the mono-energetic $\gamma$-ray sources. After optimizing the detector parameters, one can estimate the efficiencies corresponding to the unexplored energy region as well in the simulation. 

\begin{table}[]
    \centering
    \begin{tabular}{l p{1.8cm} p{0.6cm} p{0.6cm} p{0.7cm} p{1.1cm} }
    \hline
    Source     &      Reaction    &  $A_t$ ($\mu$Ci)  &  $E_b$ (MeV) & $E_\gamma$ (keV) & $T_t$ ($mg/cm^2$)\\
    \hline
   
    $^{51}$Cr     &   $^{51}$V(p,n)$^{51}$Cr &  1.445   & 10.5  &   320.08 & 2.4 \\
    $^{65}$Zn      &   $^{65}$Cu(p,n)$^{65}$Zn    & 1.724 & 10.5  & 1115.3 & 10.375\\
    $^{109}$Cd   &   $^{109}$Ag(p,n)$^{109}$Cd   & 0.556  & 9.8   & 88.033 & 4.7\\
    \hline
    
    \end{tabular}
    \caption{Fabricated monoenergetic sources, corresponding reactions, activity ($A_t$), Beam energy ($E_b$), $\gamma$-ray energy ($E_\gamma$), thickness of the target ($T_t$)}
    \label{tab:source data table}
\end{table}

\begin{figure*}
     \centering
     \begin{tabular}{cc}
         \includegraphics[scale=0.32]{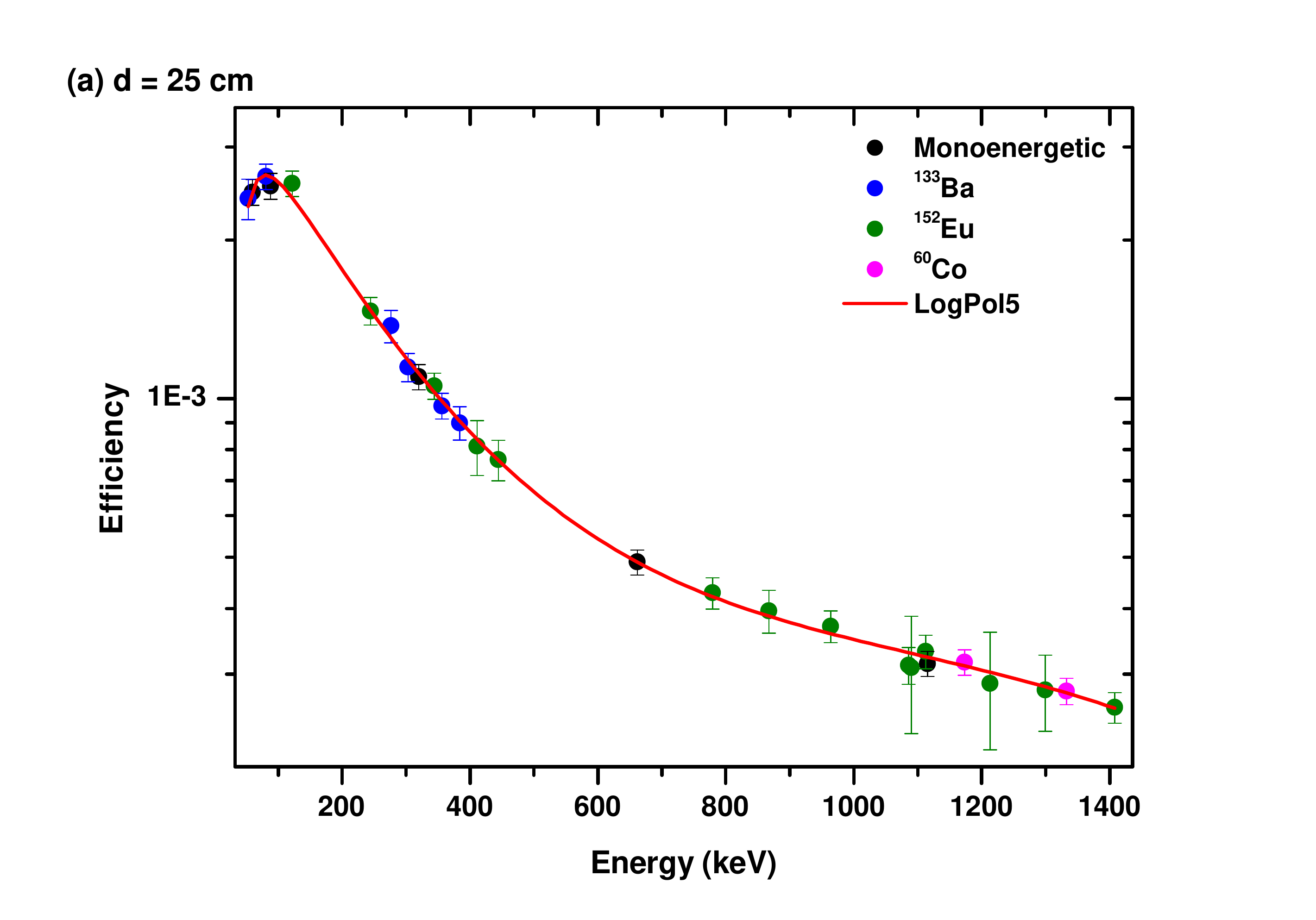}
         \includegraphics[scale=0.32]{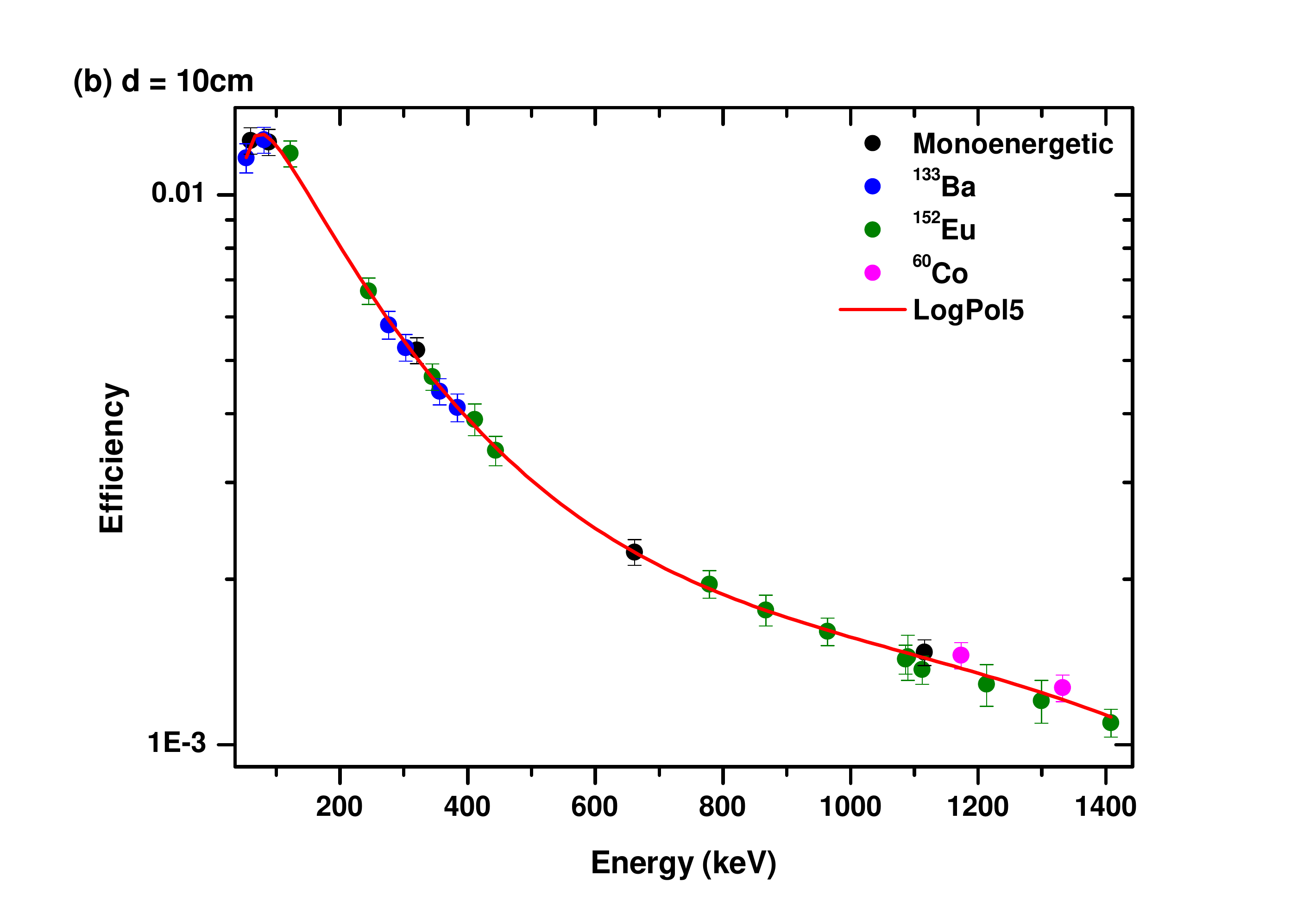}\\
         \includegraphics[scale=0.32]{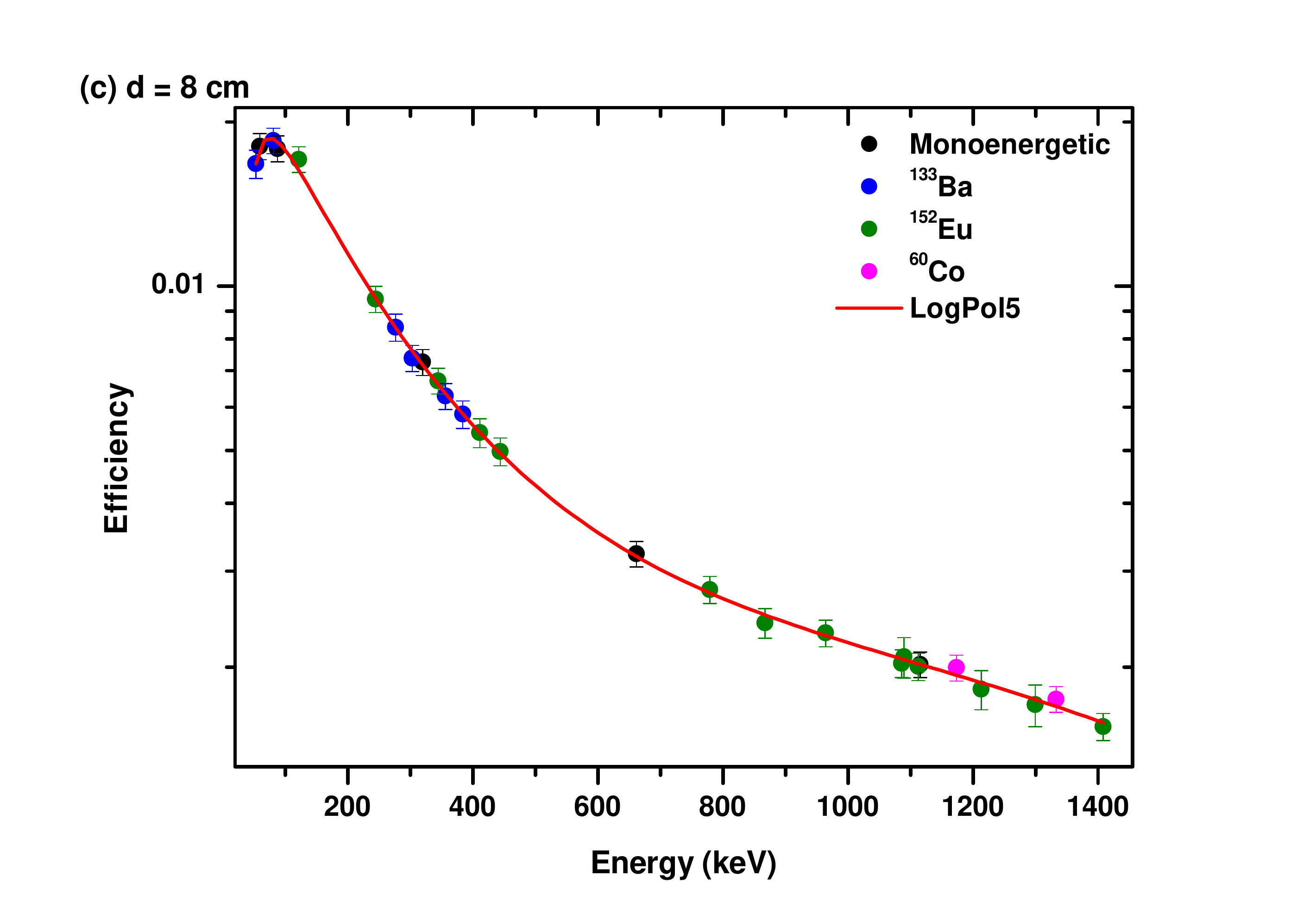}
         \includegraphics[scale=0.32]{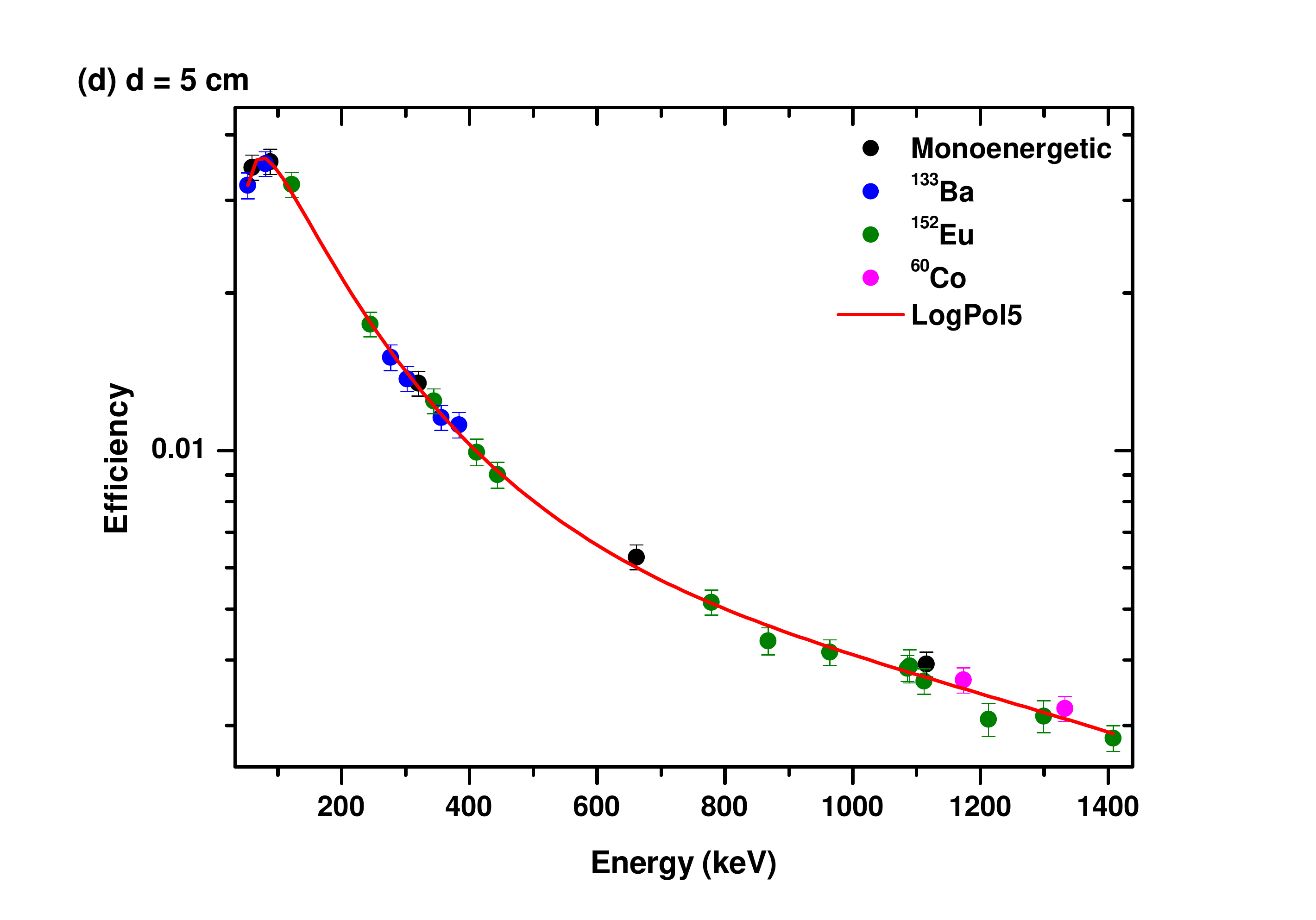}\\
         \includegraphics[scale=0.32]{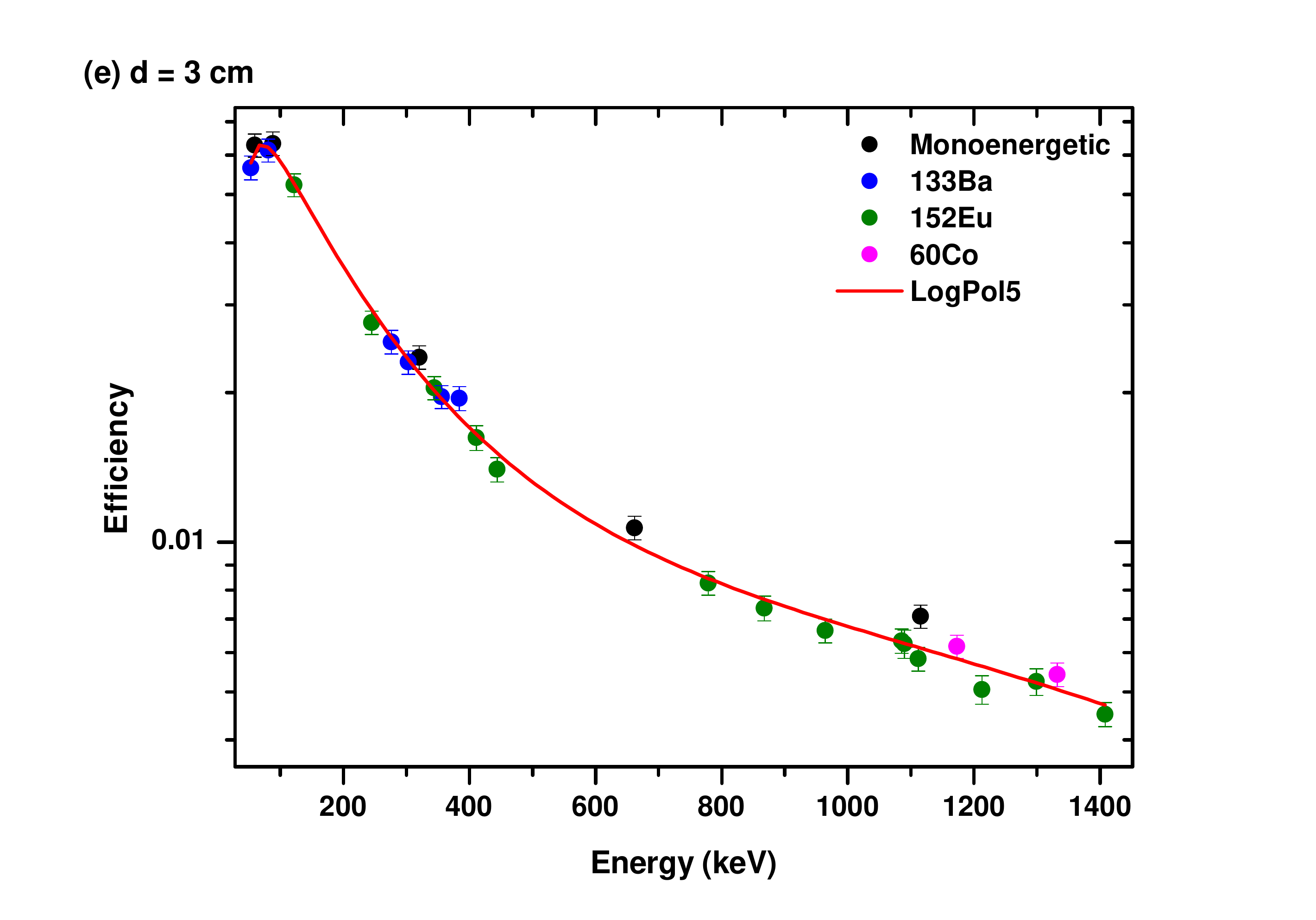} 
         \includegraphics[scale=0.32]{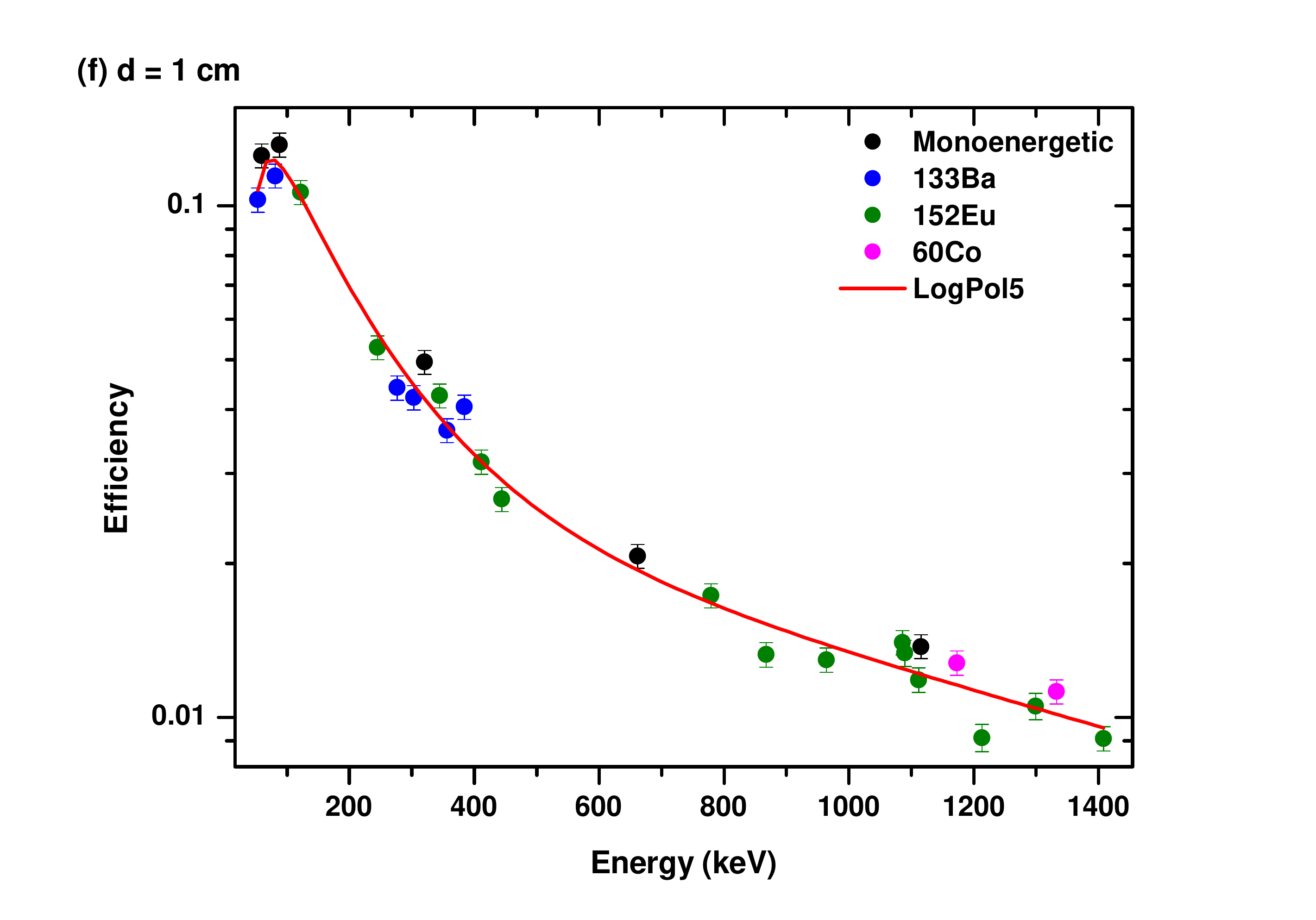}\\
    \end{tabular}
        \caption{Experimental efficiency as a function of $\gamma$-ray energy at source to detector distances, d = (a) 25 cm, (b) 10 cm, (c) 8 cm, (d) 5 cm, (e) 3 cm and (f) 1 cm. Fitting curve of fifth order log-log polynomial is depicted in red line colour.}
        \label{fig:Expeff}
\end{figure*}

The aim of this work is to obtain TCS correction factor ($k_{TCS}$) for $^{152}$Eu, $^{133}$Ba, $^{60}$Co radioactive sources using the analytical method as well as the experimental method. Probabilities for the coincidence of two or more $\gamma$-rays were calculated from the available decay schemes. The photopeak and total efficiencies were determined using Geant4 (G4) Monte Carlo simulation. The monoenergetic $\gamma$-ray sources were fabricated at  VECC, Kolkata using proton beam from the K-130 cyclotron accelerator. These sources were used for experimental validation of TCS correction factor ($k_{TCS}$) obtained analytically.  
\begin{figure*}
     \centering
        \includegraphics[trim={1.7cm 1.5cm 1.5cm 1.5cm},clip, scale=0.7]{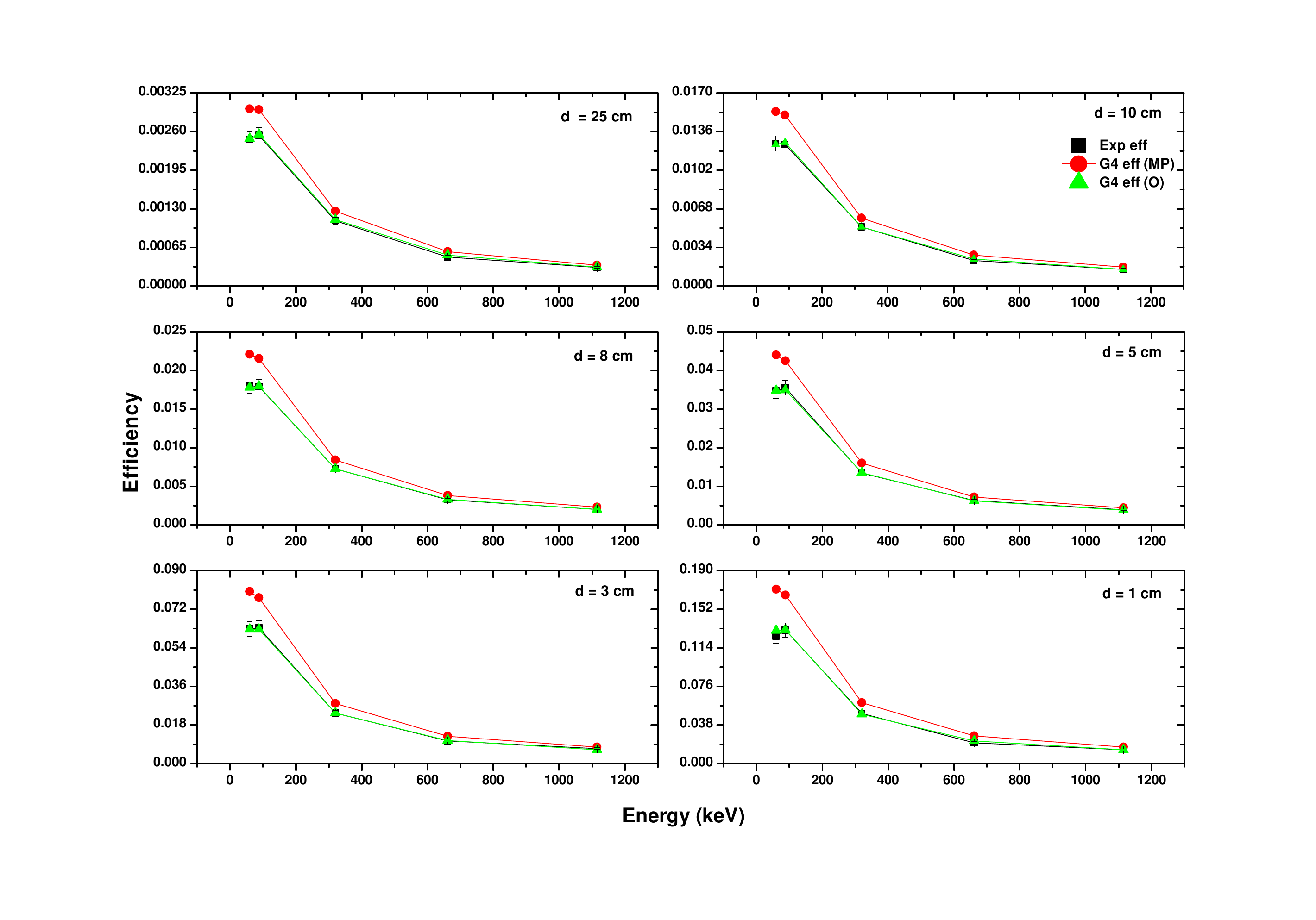}
        \caption{Efficiencies of monoenergetic sources at source to detector distances, d = 25 cm, 10 cm, 8 cm, 5 cm, 3 cm and 1 cm. Experimental efficiency shown in black colour data point, red colour data point shows the Geant4 (G4) simulated efficiency with the detector parameter provided by manufacturer (MP) and green color data point shows the Geant4 simulated efficiency with the optimized detector parameter (O).}
        \label{fig:monoenergetic}
\end{figure*}

\section{Absolute efficiency measurement using standard $\gamma$-ray sources}
\label{sec:Efficiency_measurement_using_standard_sources}
The absolute efficiency curve for an electrically cooled Falcon 5000 BEGe {\color{blue}\textbf{\cite{Abdelati2020,Sathi2020}}} detector has been determined in this work. This detector is coupled with a pre-amplifier, a buit-in MCA, Genie-2000 software to collect the spectrum and tablet connected though a LAN cable is used to control the detector functions as well as display the spectrum. The Falcon 5000 BEGe detector is a complete self reliant system with its own in-built High Voltage Power Supply (HVPS) unit, Digital Signal Processing (DSP) unit and analysis software Genie-2000. A -3500 voltage of reverse bias was applied to the detector which was supplied by the HVPS unit. The DSP unit is responsible for input pulse processing and amplification. Genie-2000 {\color{blue}\textbf{\cite{Genie}}} software acquires and analyses the data from Multichannel Analyzers (MCAs).  Genie 2000 has adequate set of capabilities to handle multiple functions like MCA control, spectral display, basic spectrum analysis and detector operations. A schematic diagram of detector geometry is shown in {\color{blue}\textbf{Fig. \ref{fig:BEGe detctor designe}}}.  The $\gamma$-ray spectrum were recorded using Genie 2000 software and it was utilized for the preliminary analysis of recorded data. The final offline analysis was performed using ROOT data analysis framework {\color{blue}\textbf{\cite{root}}} software.

The data was taken initially by placing the standard $\gamma$-ray sources at 25 and 10 cm away from the detector face {\color{blue}\textbf{\cite{Lemasson2009}}}. The calibrated standard radioactive $\gamma$-ray point sources $^{152}$Eu, $^{133}$Ba, $^{60}$Co, $^{241}$Am and $^{137}$Cs were used. The large distances of 25 cm and 10 cm between source and detector were chosen to avoid any influence of coincidence summing effect. The close geometry measurement for source to detector distance d = 8, 5, 3 and 1 cm were also performed. The experimental efficiency for a photopeak is given by,

\begin{equation}
    \epsilon = \frac{C/sec}{I_\gamma A_t}
\end{equation}

where $\epsilon, C, I_\gamma$ and $A_t$ are the efficiency of detector, area under a $\gamma$-ray photopeak, intensity of a $\gamma$-ray photopeak and activity of source, respectively {\color{blue}\textbf{\cite{MyfirstDAE}}}. The $\gamma$-ray intensities were taken from the available decay scheme in the literature {\color{blue}\textbf{\cite{recomendeddata}}}. 

 As discussed in previous section, close geometry measurements introduces summing effect in efficiency measurements. One way to avoid is to use mono-energetic $\gamma$-ray sources for efficiency measurements so that there is no cascading $\gamma$-rays to contribute in coincidence summing. However, the availability of standard mono-energetic $\gamma$-ray sources having sufficiently long half-lives is very limited and also not sufficient for broad energy region efficiency measurement. The other way to deal with the summing effect is do a TCS correction for detector efficiency measurement. This TCS correction factor can be calculated by  {\color{blue}\textbf{Eq. \ref{correcton factor}}}. To calculate the correction factor using this method we need the probability of the coincidence of $\gamma$-rays, photopeak and total efficiency of corresponding $\gamma$-ray. Total efficiency for each $\gamma$-ray photopeak was obtained by simulating the detector response at different source to detector distances using Geant4 simulation toolkit.   

\section{Geant4 Monte Carlo simulation}
\label{sec:geant4}
A Monte Carlo simulation code Geant4 {\color{blue}\textbf{\cite{Geant4}}} was used to study the response of the Falcon 5000 BEGe detector. Geant4 is a toolkit for simulating the passage and interaction of different particles through different matter {\color{blue}\textbf{\cite{Agostinelli2003}}}. It has a vast range of functionality including tracking, geometry, physics model and particle generation. The various physical models like electromagnetic, decay, optical, transportation are offered in Geant4 code. A wide variety of particles, materials, elements, geometry and energy range from 250 eV to $\sim$ TeV has been accommodated in Geant4. Until now Geant4 has been found to be very useful to perform reliable Monte Carlo simulations from low energy physics to high energy physics.

To simulate the BEGe detector response at different source to detector distances several  in-built classes in Geant4 were used. The geometry of the detector was constructed using G4VUserDetectorConstruction class, material and elements were incorporated using the in-built NIST library G4NistManager class. The volume of the Ge crystal in the detector was chosen as the only scoring volume. In principle, this scoring volume is responsible for energy deposition by the interacting $\gamma$-ray in the detector crystal and further store the deposited energy in the output data file. The structural dimensions of the detector were optimized using a well-checked procedure in order to obtain accurately the full energy peak efficiency for any energy and measurement geometry. The detector dimensions parameters provided by the manufacturer and optimized detector parameters have been tabulated in {\color{blue}\textbf{Table \ref{tab:Detector parameters}}}. A schematic diagram of the Falcon 5000 BEGe detector geometry has been shown in {\color{blue}\textbf{ Fig. \ref{fig:BEGe detctor designe}}}. For particle generation G4VUserPrimaryGeneratorAction class was used in which G4ParticleGun class was the function responsible to generate the particles. The $\gamma$-rays were generated randomly and isotropically in  all directions. In each run, 10$^8$ particles were sampled to reduce any statistical uncertainties in the simulations. Our primary particles were the $\gamma$-rays which interact  with matter through various processes, such as photo-electric effect, Compton scattering and pair production. These interactions were included in the simulation using the G4EmStandardPhysics class. The particle was tracked throughout the world volume and the energy deposited in the detector is collected in every step and later added at the end of the event. This deposited energy is stored in a root file at the end of each event and further analysis was done using ROOT data analysis framework {\color{blue}\textbf{\cite{root}}} software. The resolution of the detector was measured experimentally at different energies using photopeaks of $^{152}$Eu, $^{133}$Ba radioactive sources. The photopeak energies were fitted with Gaussian function and the corresponding standard deviations ($\sigma_{G}$) were obtained {\color{blue}\textbf{\cite{Saha2016}}}. The variation of standard deviation ($\sigma_G$) as function of energy was fitted with a function given by,

\begin{equation}\label{sigma}
    \sigma_{G} = a+b*E^{1/2}
\end{equation}

where E is energy of $\gamma$-ray in keV, a and b are the constants. The fitted value of a = 0.19259 and b = 0.0177498 were used in the simulation of detector resolution. The detector resolution was incorporated in the simulated results by redistributing the simulated spectrum with a random variable biased to a Gaussian distribution of standard deviation obtained from {\color{blue}\textbf{Eq. \ref{sigma}}}.

\begin{figure*}
     \centering
     \begin{tabular}{cc}
         \includegraphics[scale=0.32]{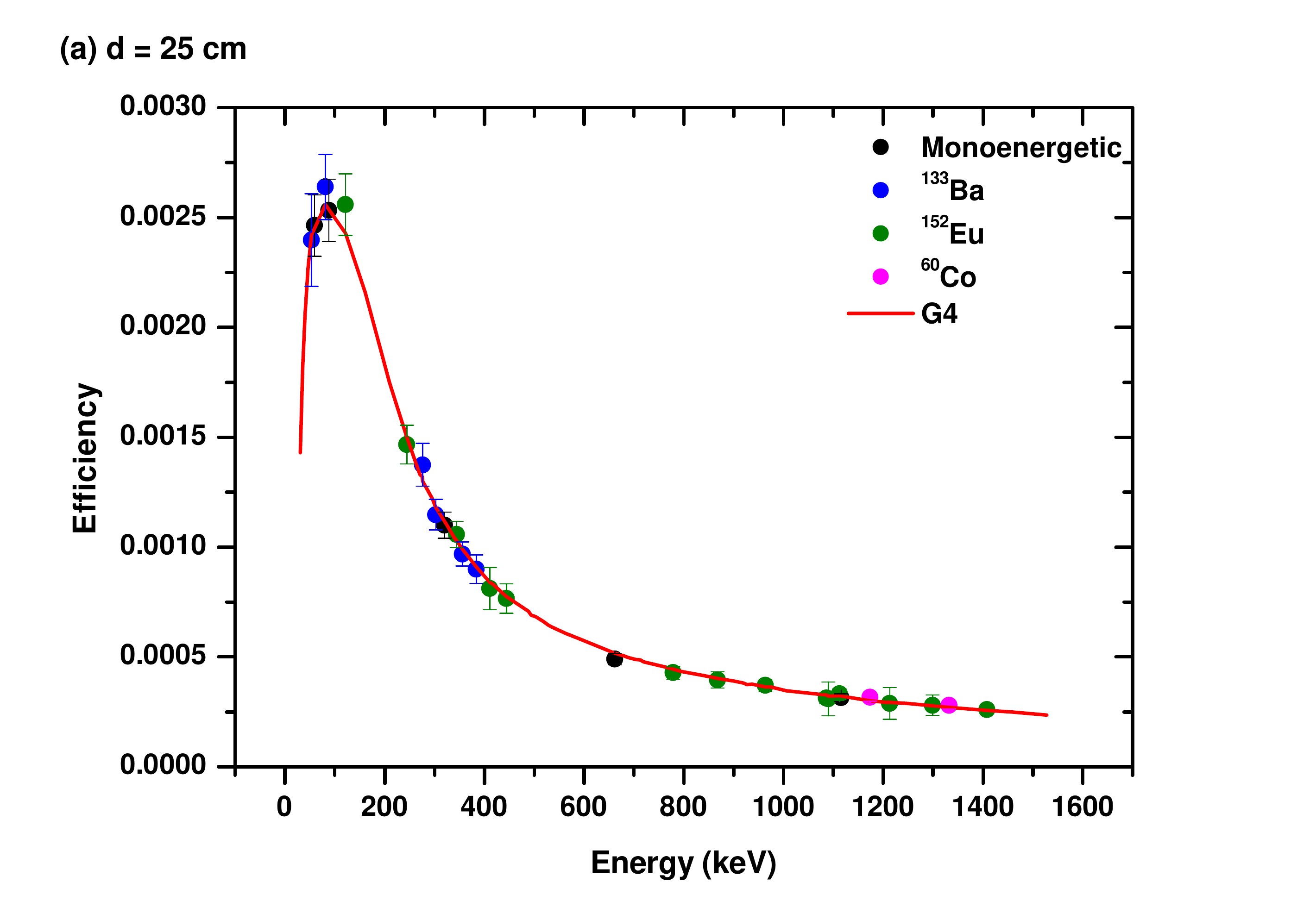}
         \includegraphics[scale=0.32]{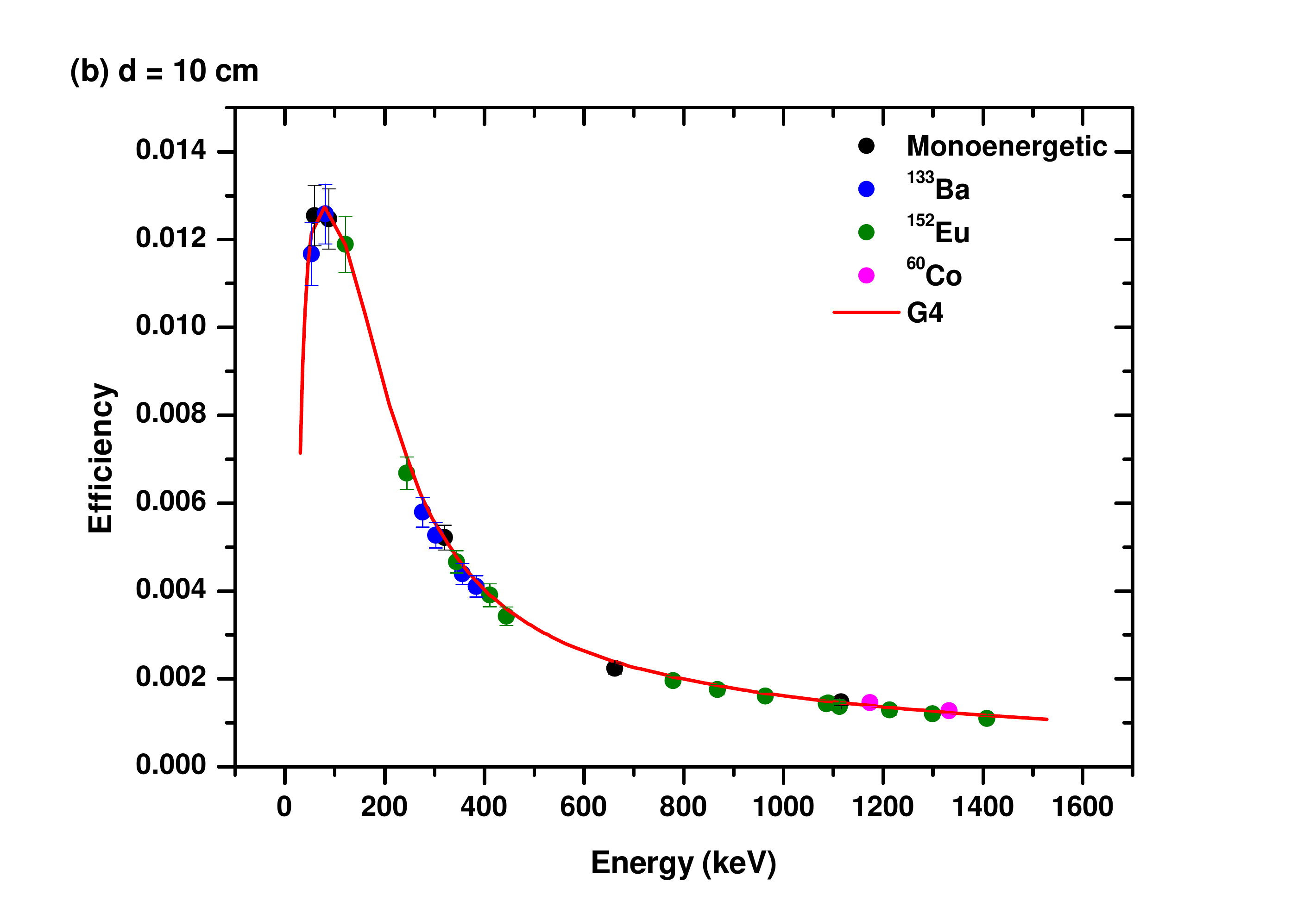}\\
         \includegraphics[scale=0.32]{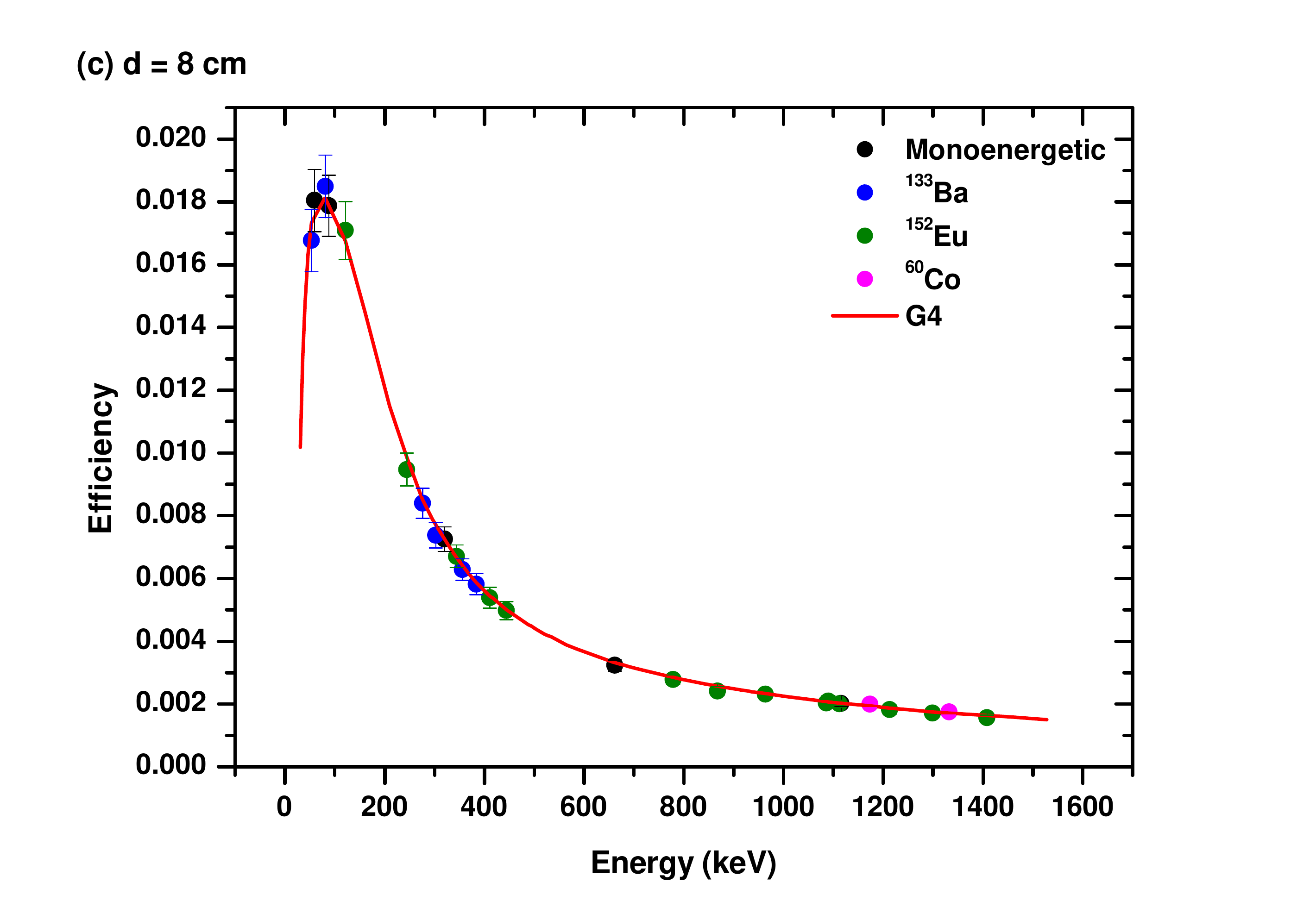}
         \includegraphics[scale=0.32]{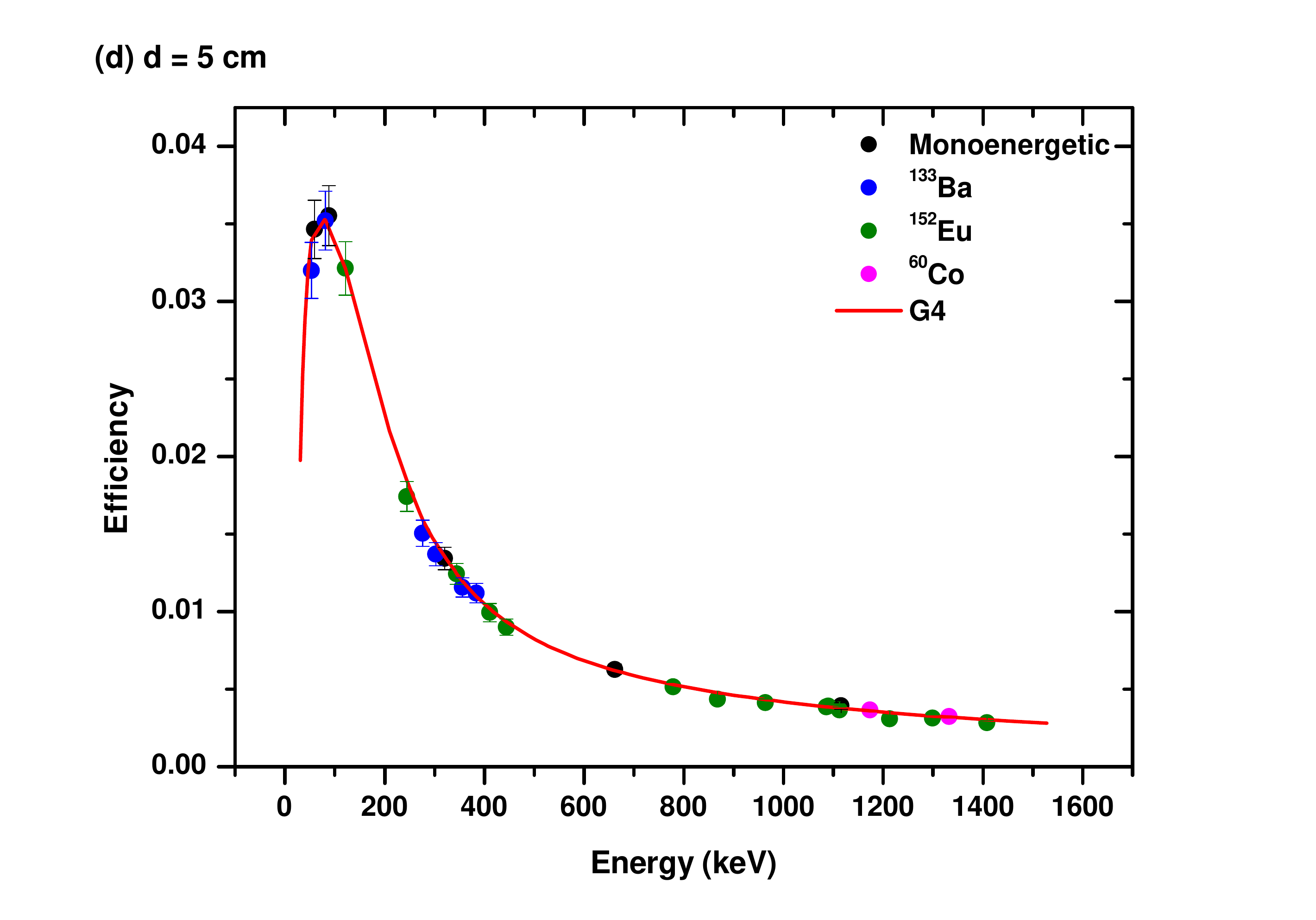}\\
         \includegraphics[scale=0.32]{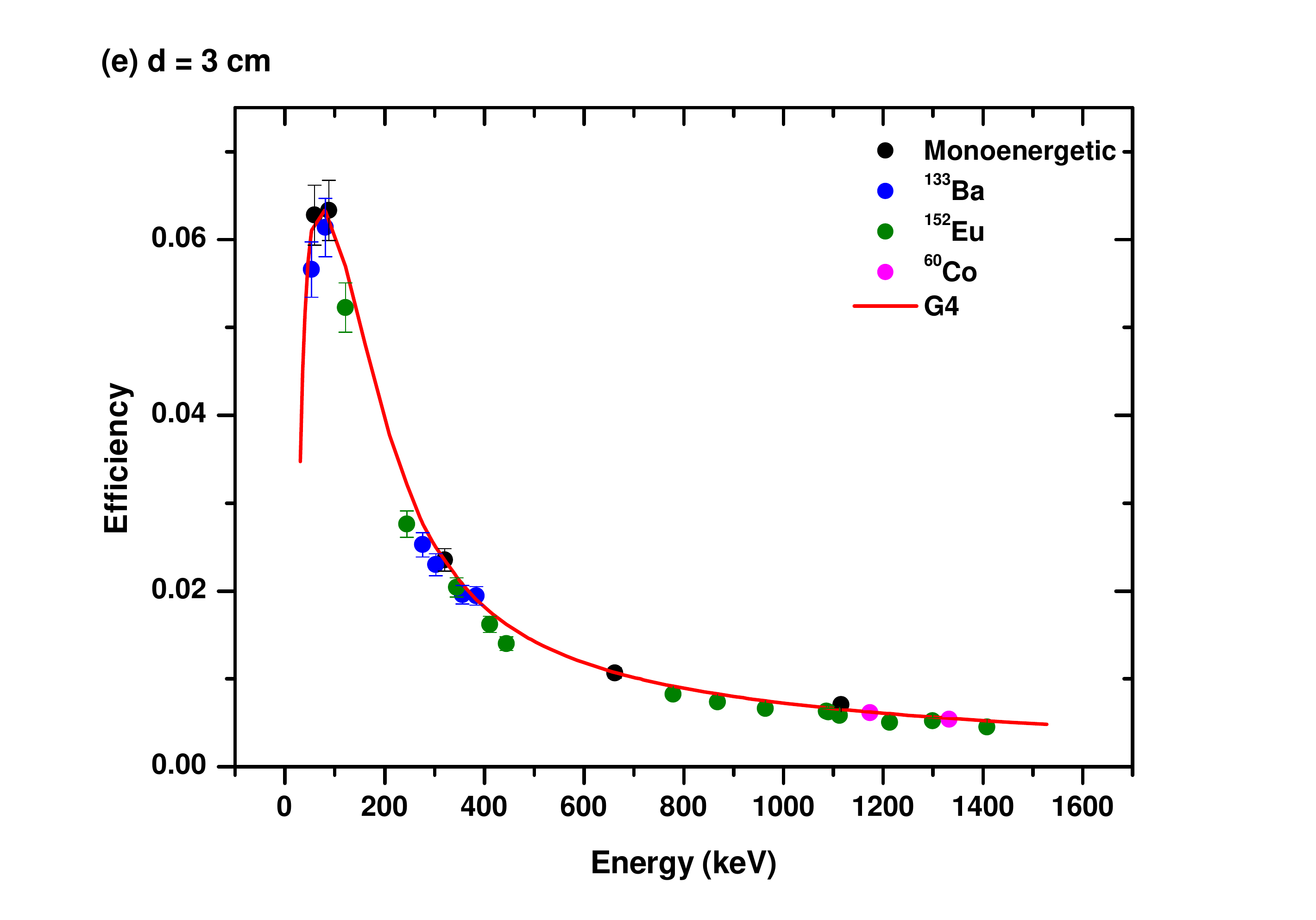}
         \includegraphics[scale=0.32]{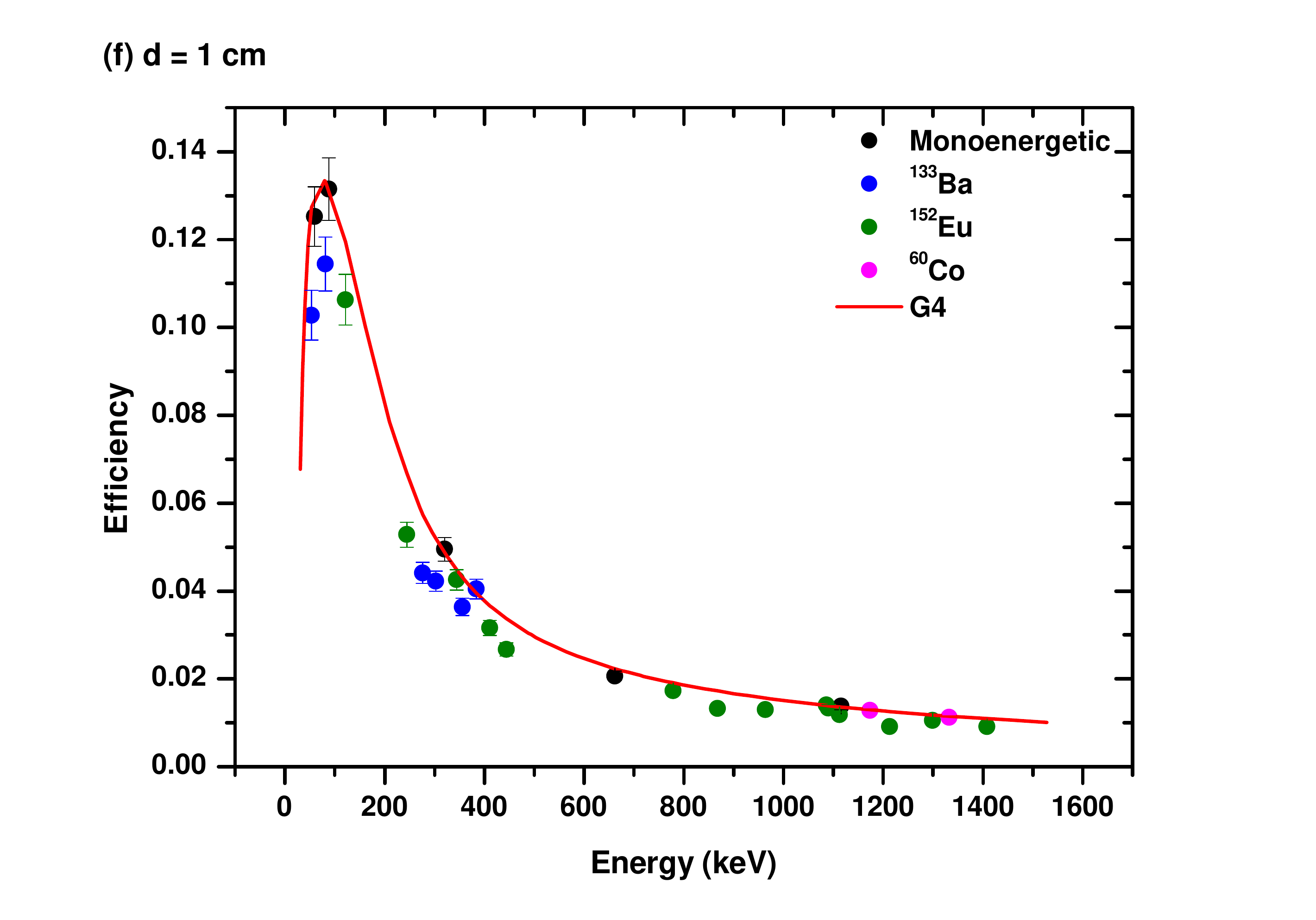}\\
    \end{tabular}
        \caption{Experimental efficiency and Geant4 (G4) simulated efficiency (in red colour line) as a function of $\gamma$-ray energy at source to detector distances, d = (a) 25 cm, (b) 10 cm, (c) 8 cm, (d) 5 cm, (e) 3 cm and (f) 1 cm.}
        \label{fig:ExpnG4}
\end{figure*} 

\section{Fabrication of mono-energetic $\gamma$-ray sources}
\label{Fabrication_of_mono-energetic_sources}
Geant4 simulation needs an experimental verification such that the simulated results are reliable in close geometry measurement. In close geometry measurement, we can not rely on standard multi-energetic sources because of the coincidence summing effect so it is required to fabricate some mono-energetic sources that are free from coincidence summing.  

\begin{table}[]
    \centering
    \begin{tabular}{l | p{0.8cm} |p{0.8cm} |p{0.9cm} |p{0.8cm} |p{0.9cm} }
\hline
E$_\gamma$ (keV) & Exp eff & G4 eff (MP) & G4/Exp (MP) & G4 eff (O) & G4/Exp (O)   \\
\hline
\underline{d = 25 cm}	&		&		&		&		&		\\ 
59.54   	&	0.0025	&	0.0030	&	1.21	&	0.0025	&	1.01	\\
88.03   	&	0.0025	&	0.0030	&	1.17	&	0.0026	&	1.01	\\
320.08  	&	0.0011	&	0.0013	&	1.15	&	0.0011	&	1.02	\\
661.66  	&	0.0005	&	0.0006	&	1.18	&	0.0005	&	1.06	\\
1115.5  	&	0.0003	&	0.0004	&	1.11	&	0.0003	&	1.01	\\
Avg. ratio  &           &           &   1.16    &           &  1.02     \\
\hline
\underline{d = 10 cm}	&		&		&		&		&		\\ 
59.54	    &	0.0125	&	0.0154	&	1.22	&	0.0125	&	0.99	\\
88.03   	&	0.0125	&	0.0151	&	1.21	&	0.0127	&	1.01	\\
320.08  	&	0.0052	&	0.0060	&	1.15	&	0.0052	&	1.00	\\
661.66	    &	0.0022	&	0.0027	&	1.22	&	0.0024	&	1.07	\\
1115.5  	&	0.0015	&	0.0016	&	1.12	&	0.0015	&	0.99	\\
Avg. ratio  &           &           &   1.18    &           &   1.01    \\
\hline
\underline{d = 8 cm}	&		&		&		&		&		\\ 
59.54   	&	0.0180	&	0.0221	&	1.22	&	0.0178	&	0.99	\\
88.03   	&	0.0179	&	0.0216	&	1.21	&	0.0180	&	1.01	\\
320.08   	&	0.0073	&	0.0084	&	1.16	&	0.0072	&	1.00	\\
661.66	    &	0.0032	&	0.0038	&	1.18	&	0.0033	&	1.03	\\
1115.5  	&	0.0020	&	0.0023	&	1.14	&	0.0020	&	1.00	\\
Avg. ratio  &           &           &   1.18    &           &   1.00    \\
\hline
\underline{d = 5 cm}	&		&		&		&		&		\\ 
59.54	    &	0.0347	&	0.0440	&	1.27	&	0.0348	&	1.01	\\
88.03   	&	0.0355	&	0.0425	&	1.20	&	0.0349	&	0.98	\\
320.08  	&	0.0134	&	0.0160	&	1.19	&	0.0135	&	1.01	\\
661.66	    &	0.0063	&	0.0072	&	1.15	&	0.0062	&	0.99	\\
1115.5	    &	0.0039	&	0.0044	&	1.11	&	0.0038	&	0.96	\\
Avg. ratio  &           &           &   1.18    &           &   0.99    \\
\hline
\underline{d = 3 cm}	&		&		&		&		&		\\ 
59.54   	&	0.0628	&	0.0803	&	1.28	&	0.0627	&	1.00	\\
88.03   	&	0.0633	&	0.0774	&	1.22	&	0.0626	&	0.99	\\
320.08  	&	0.0235	&	0.0281	&	1.20	&	0.0235	&	1.00	\\
661.66  	&	0.0107	&	0.0127	&	1.19	&	0.0107	&	1.01	\\
1115.5  	&	0.0071	&	0.0077	&	1.08	&	0.0065	&	0.92	\\
Avg. ratio  &           &           &   1.19    &           &   0.98    \\
\hline
\underline{d = 1 cm}	&		&		&		&		&		\\ 
59.54   	&	0.1253	&	0.1716	&	1.37	&	0.1313	&	1.05	\\
88.03   	&	0.1315	&	0.1661	&	1.26	&	0.1319	&	1.00	\\
320.08  	&	0.0495	&	0.0599	&	1.21	&	0.0488	&	0.99	\\
661.66	    &	0.0207	&	0.0272	&	1.32	&	0.0224	&	1.08	\\
1115.5  	&	0.0138	&	0.0164	&	1.19	&	0.0136	&	0.99	\\
Avg. ratio  &           &           &   1.27    &           &   1.02    \\
\hline

    \end{tabular}
    \caption{The comparison of experimental efficiency (Exp eff), Geant4 efficiency (G4 eff) with manufacturer provided dimension (MP) and optimized parameters (O) at source to detector distance, d = 25, 10, 8, 5, 3 and 1 cm}
    \label{tab:monoenergetic_optimization}
\end{table}

\subsection{Experiment}
\label{Experiment}
A few mono-energetic $\gamma$-ray sources were fabricated via irradiation method, using proton beams obtained from the K-130 cyclotron at VECC, Kolkata {\color{blue}\textbf{\cite{Bhattacharya2018}}}. The accelerated p beams were used to bombard the natural $^{51}$V, $^{65}$Cu and $^{109}$Ag targets to fabricate the mono-energetic $\gamma$-ray sources $^{51}$Cr, $^{65}$Zn and $^{109}$Cd. The targets were mounted perpendicular to the beam direction and  placed in beam line with aluminum foil and an aluminium (Al) plate of 1 mm thickness, which prevents the end flange getting activated on prolonged use. This end flange was being cooled by low conductivity water (LCW) to avoid any heating of target material during long irradiation runs. The duration of irradiation of each target varied from 8 hrs to 60 hrs depending on the reaction specifications such as target thickness, reaction cross-section and desired $\gamma$-ray yield. The vacuum inside the irradiation setup was between $10^{-5}-10^{-6}$ mbar through out the experiment {\color{blue}\textbf{\cite{Dasgupta}}}. A schematic diagram of the experimental setup is shown in {\color{blue}\textbf{Fig. \ref{fig:accelerator ecperimental setup}}}. The beam current was around 400-500 nA during the course of irradiation. Beam was dumped on Al plate and  beam current was measured directly from the end flange using a current integrator. The details of the monoenergetic radioactive sources, corresponding reactions, beam energy ($E_b$), thickness of the target material ($T_t$), observed the $\gamma$-ray ($E_\gamma$) after irradiation of target and calculated activity ($A_t$) has been tabulated in the {\color{blue}\textbf{ Table \ref{tab:source data table}}}.

\subsection{Activity measurement of the fabricated mono-energetic $\gamma$-ray sources}
\label{Activity_measurement_of_mono-energetic_sources}
The measured efficiency curve obtained at distances of 25 cm and 10 cm using standard  $^{152}$Eu, $^{133}$Ba, $^{60}$Co and $^{137}$Cs point radioactive sources were used to standardize the fabricated mono-energetic $\gamma$-ray sources. The efficiencies of point sources at these large distance were fitted with log-log polynomial function of fifth order. Activities (disintegration rates) of fabricated source were calculated by interpolating the efficiencies from these curves and average of the activities at two distances were chosen as the actual activity of the fabricated sources. The fabricated standard mono-energetic sources thus obtained were also used for the efficiency measurements at close geometries where the source to detector distances were d = 8, 5, 3, 1 cm, respectively. These measurements were used in validation of Geant4 simulation. 

\begin{table*}[]
    \centering
    \begin{tabular}{l m{1cm} | p{0.8cm} p{0.7cm} | p{0.8cm} p{0.7cm} | p{0.8cm} p{0.7cm} | p{0.8cm} p{0.7cm} | p{0.9cm} p{0.7cm} | p{0.9cm} p{0.7cm}}
   
   \hline
    Source & E$_\gamma$ (keV) & $k^{Analytical}_{TCS}$ & $k^{Exp}_{TCS}$ &  $k^{Analytical}_{TCS}$ & $k^{Exp}_{TCS}$ & $k^{Analytical}_{TCS}$ & $k^{Exp}_{TCS}$ & $k^{Analytical}_{TCS}$ & $k^{Exp}_{TCS}$ &  $k^{Analytical}_{TCS}$ & $k^{Exp}_{TCS}$ & $k^{Analytical}_{TCS}$ & $k^{Exp}_{TCS}$   \\
    \hline
    	 &          		& 	d = 1 cm 	   &           		& 	d = 3 cm  	&           		& 	 d = 5 cm  	&		&	d = 8 cm	&		&	d = 10 cm	&		&	d = 25 cm	&		\\
    $^{60}$Co	&	1173.2	&	1.074	&	1.029	&	1.035	&	1.009	&	1.003	&	0.981	&	1.01	&	0.971	&	1.007	&	0.96	&	1.001	&	0.949	\\
	&	1332.5	&	1.077	&	1.042	&	1.036	&	1.023	&	1.003	&	0.988	&	1.011	&	0.981	&	1.008	&	0.97	&	1.001	&	0.969	\\
    $^{133}$Ba	&	53.16	&	1.208	&	1.239	&	1.091	&	1.078	&	1.05	&	1.061	&	1.026	&	1.034	&	1.018	&	1.041	&	1.001	&	1.008	\\
	&	80.99	&	1.13	&	1.166	&	1.059	&	1.03	&	1.033	&	1.001	&	1.017	&	0.98	&	1.012	&	1.011	&	1	&	0.968	\\
	&	276.39	&	1.277	&	1.301	&	1.118	&	1.094	&	1.063	&	1.058	&	1.032	&	1.013	&	1.022	&	1.053	&	1.003	&	0.947	\\
	&	302.85	&	1.176	&	1.23	&	1.079	&	1.086	&	1.043	&	1.051	&	1.022	&	1.045	&	1.015	&	1.047	&	1.002	&	1.031	\\
	&	356.01	&	1.157	&	1.186	&	1.071	&	1.06	&	1.039	&	1.034	&	1.02	&	1.022	&	1.014	&	1.047	&	1.001	&	1.022	\\
	&	383.85	&	0.938	&	0.977	&	0.969	&	0.976	&	0.982	&	0.979	&	0.991	&	1.007	&	0.993	&	1.028	&	0.998	&	1.011	\\
    $^{152}$Eu	&	121.78	&	1.08	&	1.124	&	1.037	&	1.09	&	1.021	&	0.999	&	1.011	&	0.98	&	1.008	&	0.998	&	1.002	&	0.95	\\
	&	244.69	&	1.17	&	1.225	&	1.075	&	1.134	&	1.04	&	1.059	&	1.021	&	1.041	&	1.015	&	1.054	&	1.003	&	1.023	\\
	&	344.27	&	1.066	&	1.054	&	1.031	&	1.058	&	1.017	&	0.999	&	1.009	&	0.995	&	1.007	&	1.022	&	1.001	&	0.978	\\
	&	411.11	&	1.176	&	1.159	&	1.078	&	1.087	&	1.044	&	1.024	&	1.023	&	1.009	&	1.016	&	0.998	&	1.003	&	1.034	\\
	&	443.96	&	1.29	&	1.262	&	1.122	&	1.158	&	1.065	&	1.039	&	1.033	&	1.006	&	1.023	&	1.047	&	1.004	&	1.011	\\
	&	778.9	&	1.103	&	1.104	&	1.047	&	1.109	&	1.027	&	1.028	&	1.014	&	1.025	&	1.01	&	1.043	&	1.002	&	1.036	\\
	&	867.37	&	1.255	&	1.301	&	1.108	&	1.124	&	1.058	&	1.098	&	1.029	&	1.065	&	1.02	&	1.054	&	1.004	&	1.017	\\
	&	964.07	&	1.227	&	1.205	&	1.098	&	1.132	&	1.052	&	1.047	&	1.026	&	1.008	&	1.018	&	1.038	&	1.003	&	0.985	\\
	&	1085.86	&	1.008	&	0.993	&	1.002	&	1.06	&	1	&	1	&	1	&	1.024	&	1	&	1.044	&	1	&	1.044	\\
	&	1089.73	&	1.101	&	1.041	&	1.046	&	1.067	&	1.026	&	0.988	&	1.014	&	0.987	&	1.01	&	1.027	&	1.002	&	1.044	\\
	&	1112.07	&	1.203	&	1.152	&	1.087	&	1.123	&	1.046	&	1.039	&	1.023	&	1.011	&	1.016	&	1.067	&	1.003	&	0.972	\\
	&	1212.94	&	1.418	&	1.377	&	1.165	&	1.197	&	1.087	&	1.13	&	1.043	&	1.029	&	1.03	&	1.048	&	1.006	&	1.015	\\
	&	1299.14	&	1.104	&	1.122	&	1.047	&	1.079	&	1.029	&	1.046	&	1.014	&	1.027	&	1.01	&	1.055	&	1.002	&	0.989	\\
	&	1408	&	1.22	&	1.203	&	1.094	&	1.162	&	1.051	&	1.07	&	1.025	&	1.047	&	1.018	&	1.065	&	1.003	&	0.984	\\

   \hline
   \end{tabular}
    \caption{Coincidence summing correction factors using analytical ($k^{Analytical}_{TCS}$) and experimental ($k^{Exp}_{TCS}$) method at source to detector distances, d = 1, 3, 5, 8, 10 and 25 cm }
    \label{tab:correctionfactor}
\end{table*}

\section{Results and discussion}
\label{sec:resultanddiscussion}
The measured efficiencies for the source to detector distances d = 25, 10, 8, 5, 3 and 1 cm are shown in {\color{blue}\textbf{Figs. \ref{fig:Expeff}(a-f)}} respectively. The experimental efficiencies were determined using both the standard and fabricated mono-energetic sources like $^{241}$Am, $^{109}$Cd, $^{51}$Cr, $^{137}$Cs and $^{65}$Zn. The standard radioactive mono-energetic sources $^{241}$Am and $^{137}$Cs were available in the laboratory, whereas $^{109}$Cd, $^{51}$Cr, $^{65}$Zn mono-energetic radioactive sources were prepared at K-130 cyclotron facility at  VECC, Kolkata. Experimental efficiencies were also determined using the multi-energetic sources $^{60}$Co, $^{133}$Ba and $^{152}$Eu. In the close geometry measurements, the experimental efficiencies determined from mono-energetic sources are free from TCS whereas the experimental efficiencies from multi-energetic sources will not be free from the TCS. The sources with complex decay scheme may have larger summing effect than the sources with simple decay scheme. 

These experimental efficiency curves were fitted using fifth order log-log polynomial at each source to detector distances, which is shown using solid red line in {\color{blue}\textbf{Figs. \ref{fig:Expeff} (a-f)}}. It is observed from the experimental efficiency curve that the efficiency of BEGe detector and fitting curve matches well with each other for the source to detector distances from d = 25 up to 8 cm. However the measured efficiencies are found to be deviating from the fit curve as the source to detector distance decreases further. {\color{blue}\textbf{Fig. \ref{fig:Expeff}(f)}} shows the experimental efficiencies corresponding to the source to detector distance of 1 cm and its deviation from the fit curve which clearly depicts the presence of coincidence summing effect at close geometry measurements. 

\begin{figure*}
     \centering
     \begin{tabular}{cc}
         \includegraphics[scale=0.32]{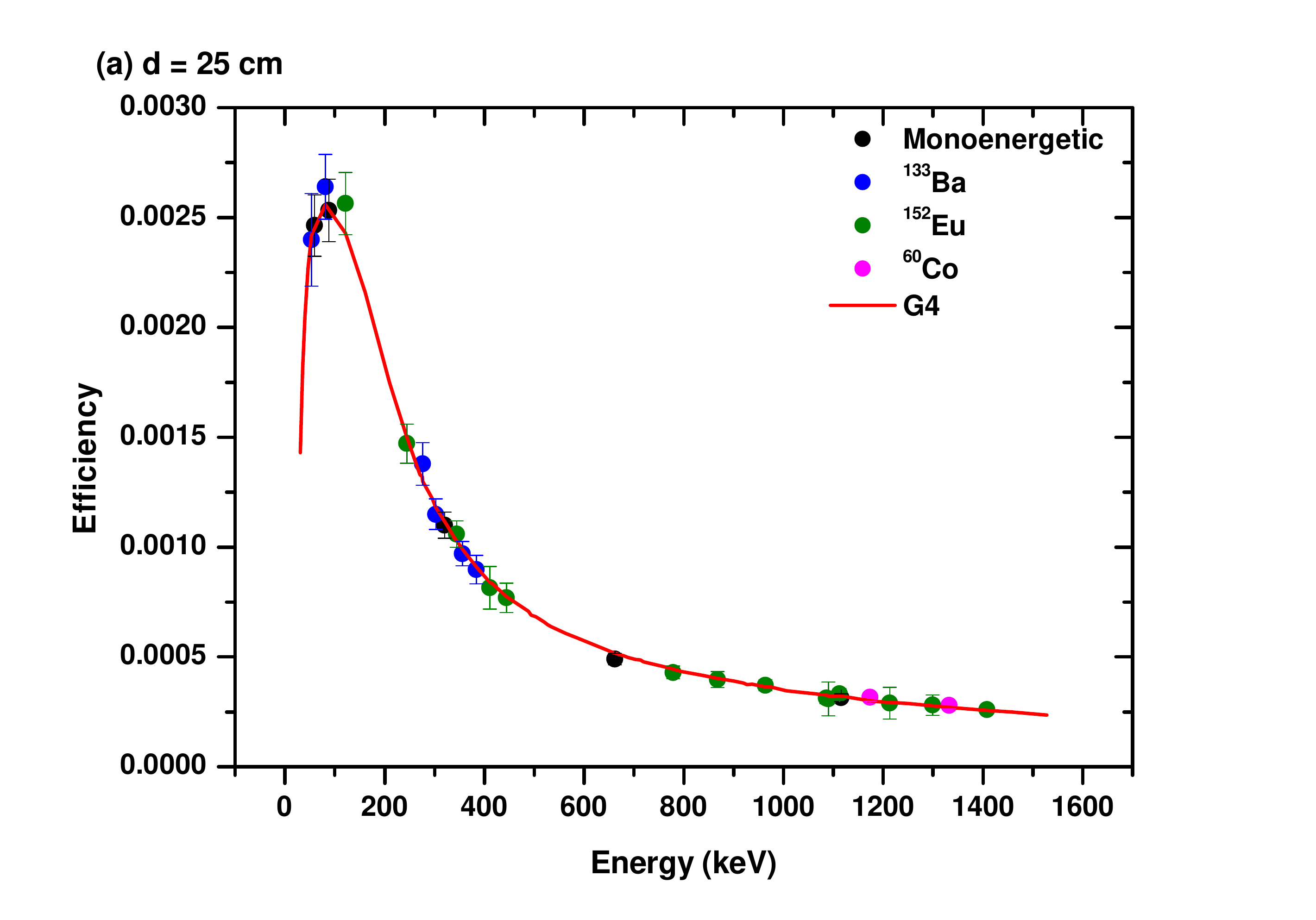}
         \includegraphics[scale=0.32]{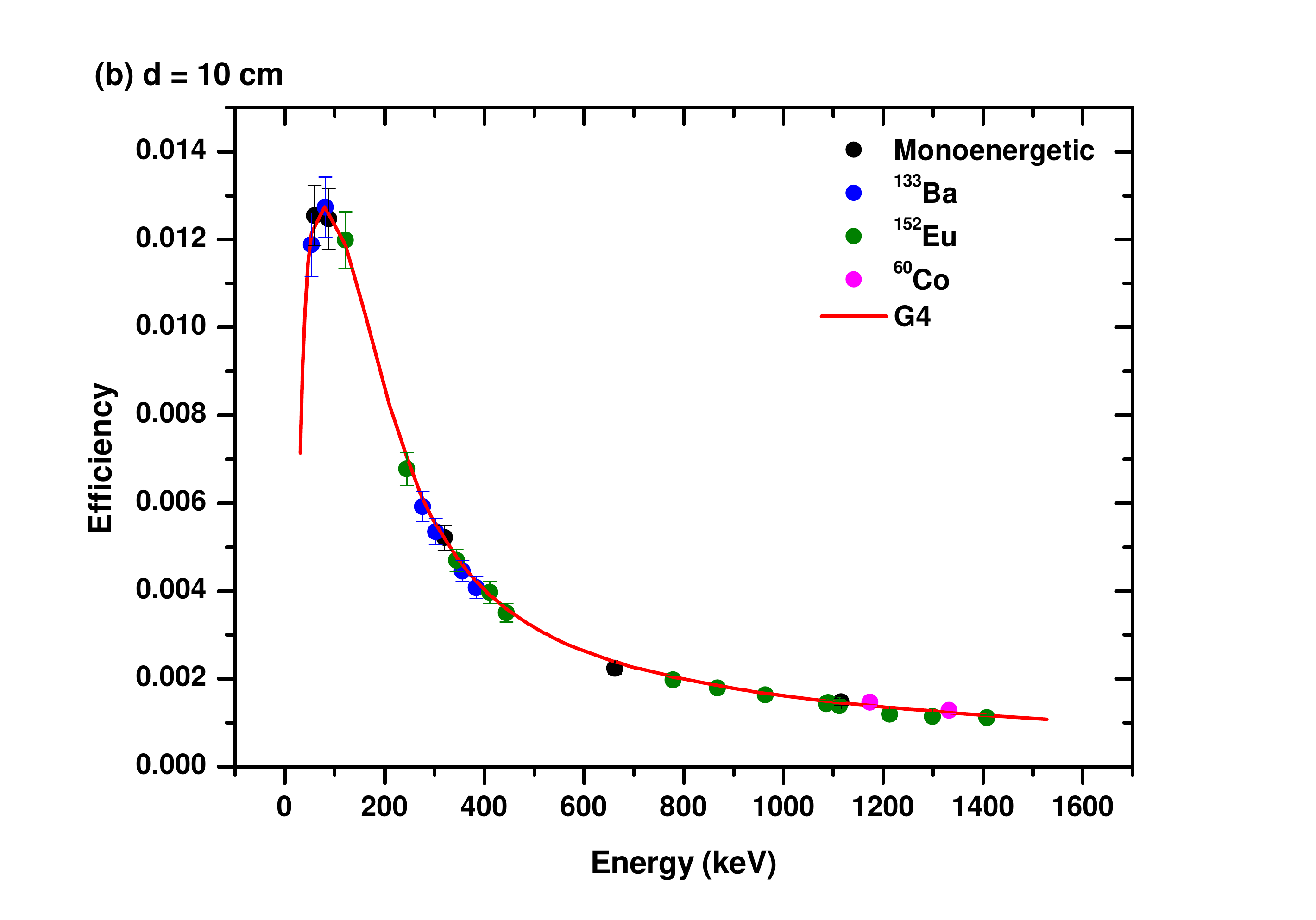}\\
         \includegraphics[scale=0.32]{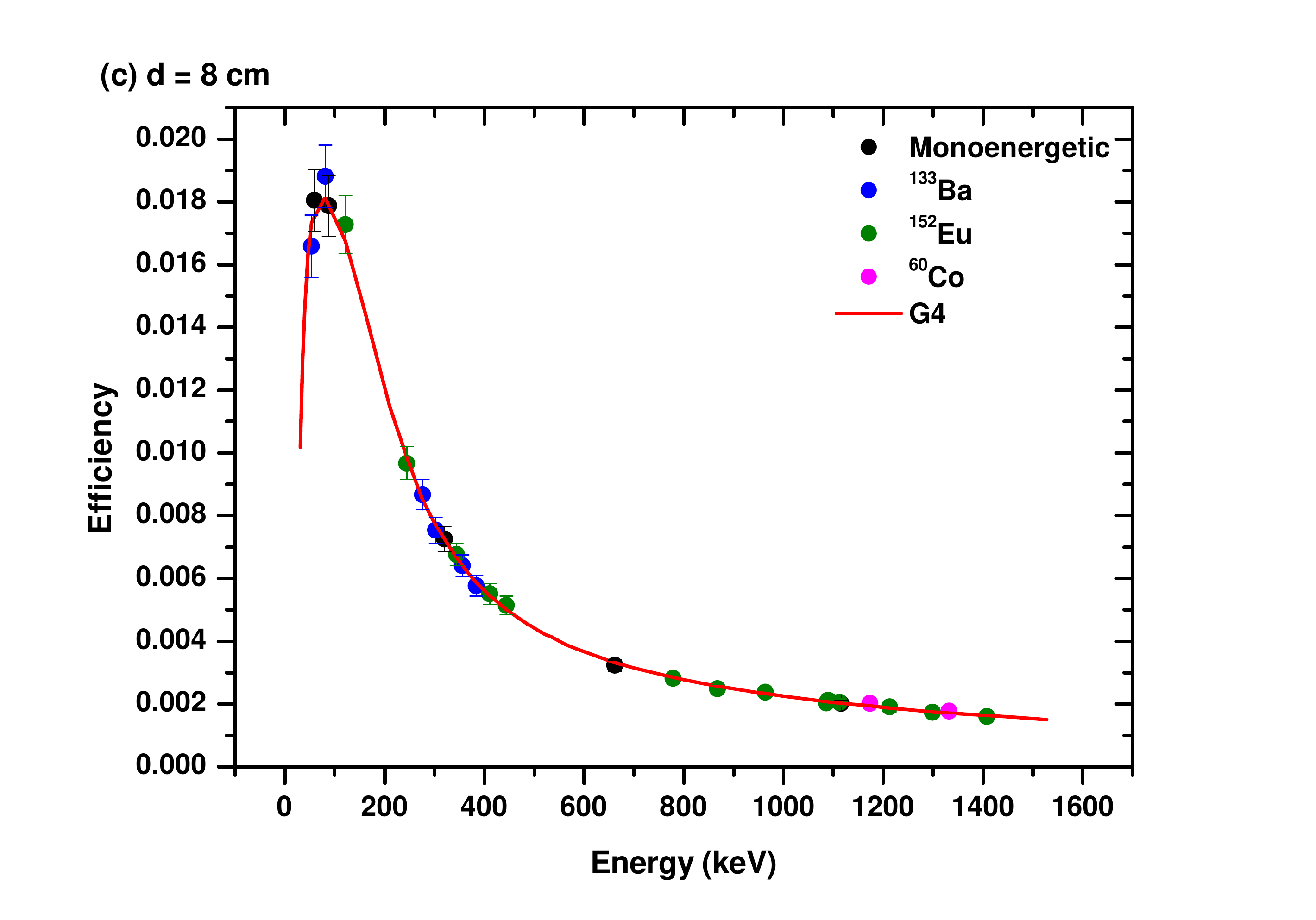}
         \includegraphics[scale=0.32]{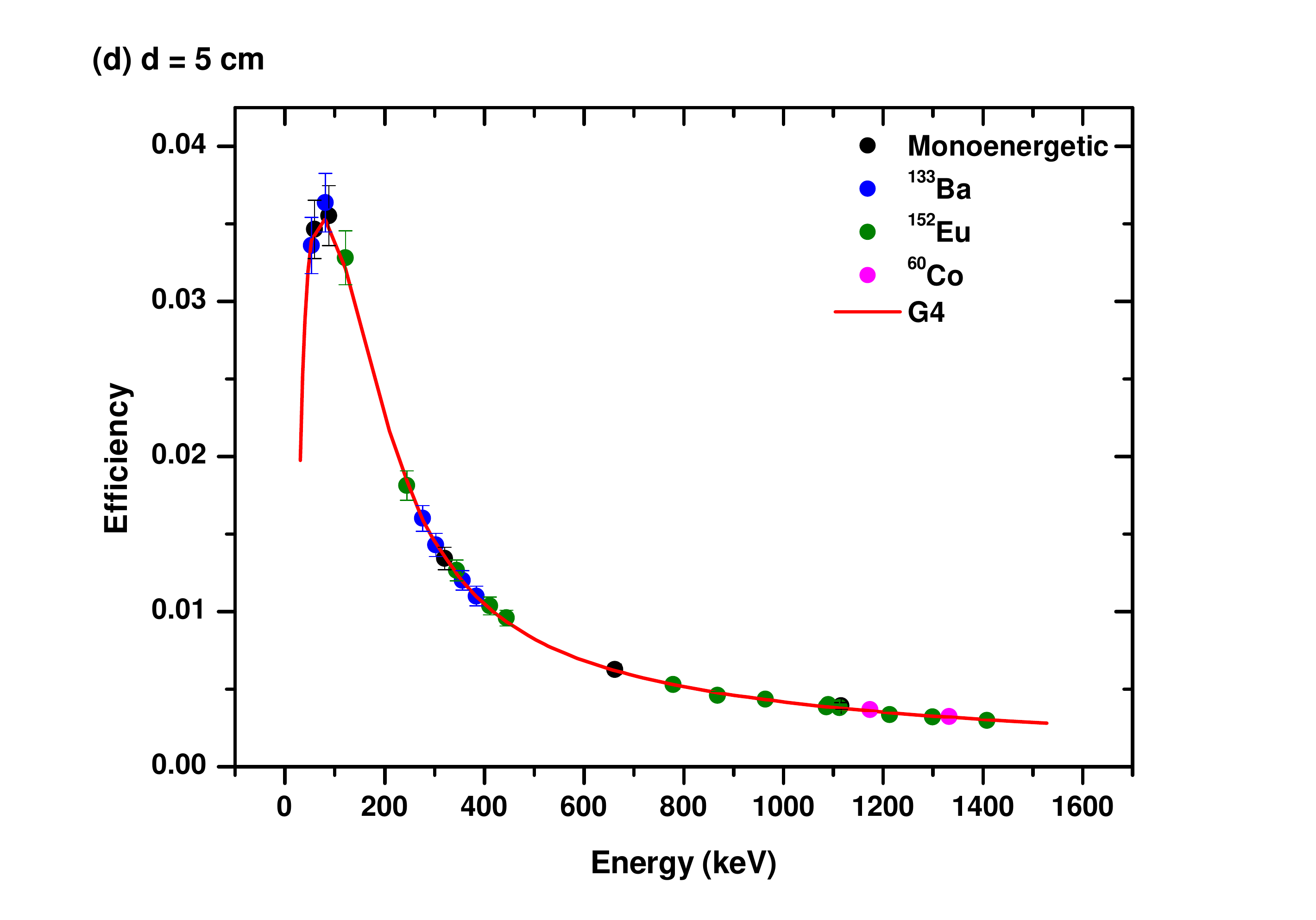}\\
         \includegraphics[scale=0.32]{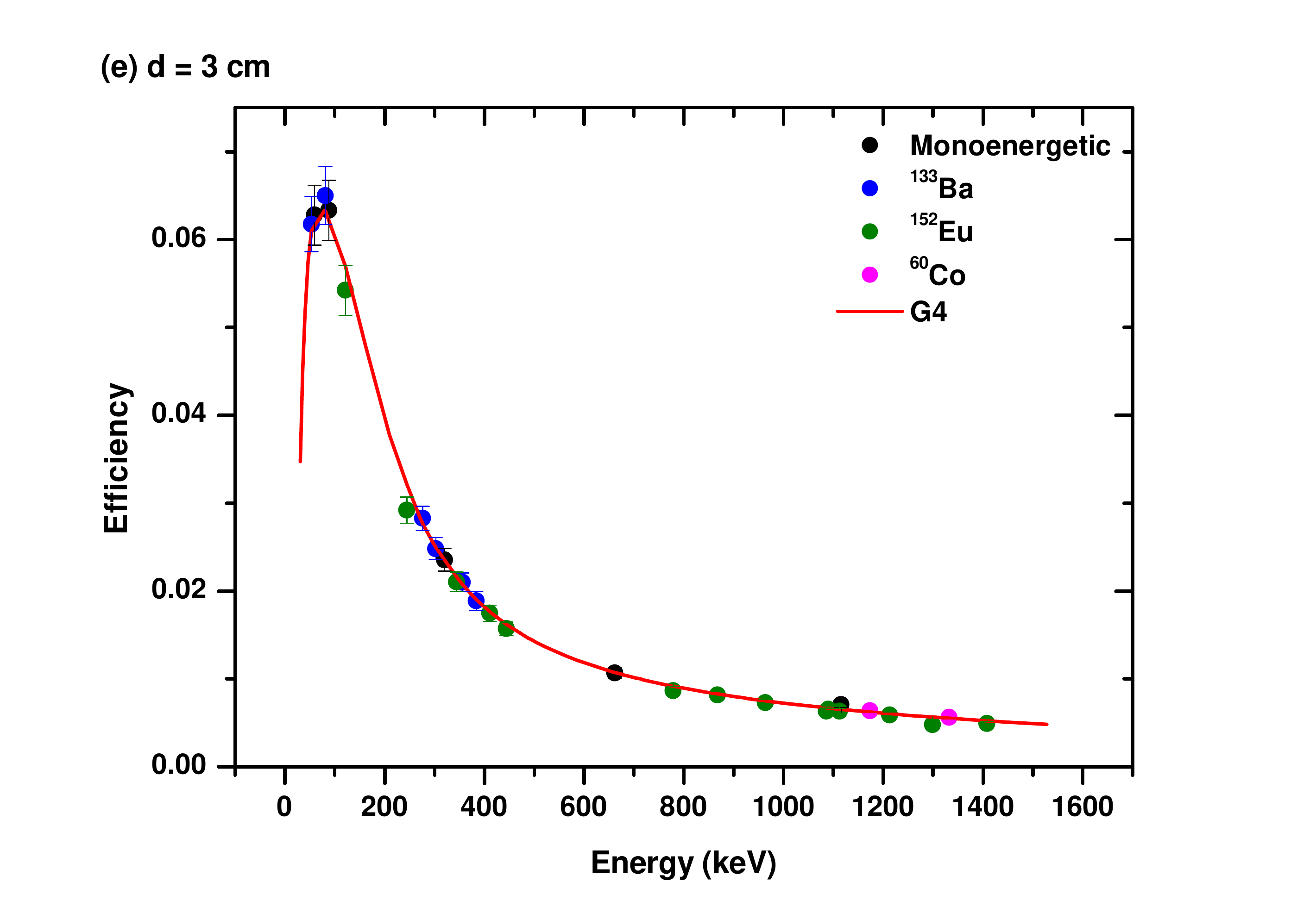}
         \includegraphics[scale=0.32]{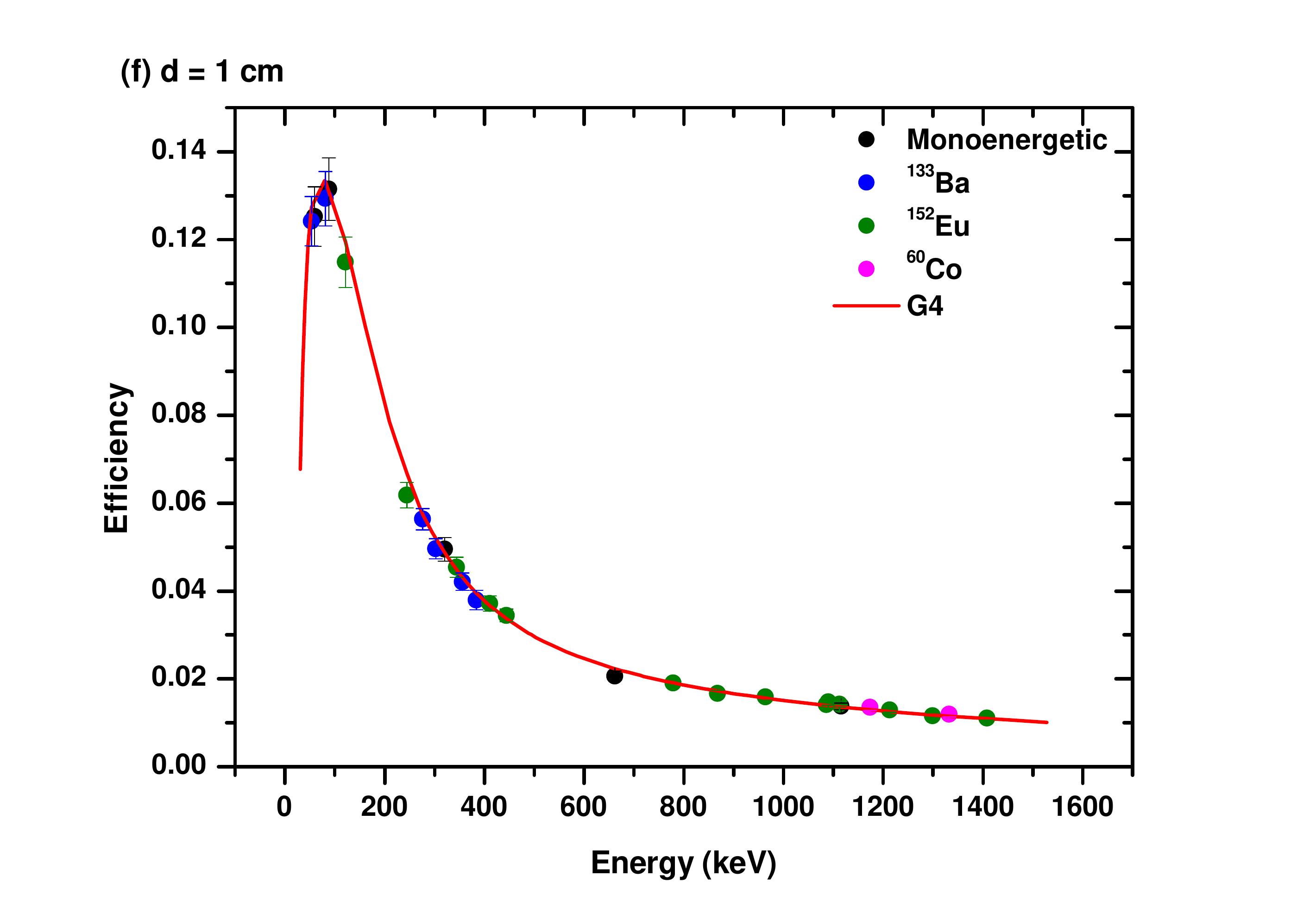}\\
    \end{tabular}
        \caption{TCS corrected experimental efficiency and Geant4 (G4) simulated efficiency (in red line colour) as function of $\gamma$-ray energy at source to detector distances, d = (a) 25 cm, (b) 10 cm, (c) 8 cm, (d) 5 cm, (e) 3 cm and (f) 1 cm.}
        \label{fig:AnanG4}
\end{figure*}

Efficiency of the detector at close geometry without the effect of coincidence summing may be determined using mono-energetic radioactive sources having desired $\gamma$-ray energy, but there are limitations with these sources as mentioned earlier. On the other hand, we can use multi-energetic $\gamma$-ray sources and calculate the coincidence summing correction factor for each $\gamma$-ray energy and determine the sum corrected efficiency. As discussed in {\color{blue}\textbf{Section \ref{sec:Introduction}}}, to calculate the coincidence summing correction factor for any $\gamma$-ray energy we need to perform the simulation to obtain the BEGe detector response for each $\gamma$-ray energy at each source to detector distance (d). Initially, the simulations were performed for all five monoenergetic sources at source to detector distances d =  25, 10, 8, 5, 3 and 1 cm, using the dimension of detector as provided by manufacturer. {\color{blue}\textbf{Fig. \ref{fig:monoenergetic}}} shows the experimental efficiency, Geant4 simulated efficiency with detector dimension provided by manufacturer (MP) and optimized detector dimension (O) by Geant4 Monte Carlo simulation for monoenergetic radioactive sources. It is evident from the {\color{blue}\textbf{Fig. \ref{fig:monoenergetic}}} that the simulated efficiencies (with manufacturer provided detector dimension) are higher than the experimental efficiency. The similar trend in efficiency were also reported in previous measurements {\color{blue}\textbf{\cite{Chhavi2011,Laborie2000,Vargas2002,Budjas2009}}}. The $\gamma$-ray energies emitted by each mono-energetic source used in this study lie in the range between 59-1115 keV. The average ratio of Geant4 (MP) to experimental efficiency for d =  25 is 1.164 and it increases up to 1.27 at d = 1 cm. The average ratio for $\gamma$-rays of mono-energetic sources at each source to detector distance has been tabulated in {\color{blue}\textbf{Table \ref{tab:monoenergetic_optimization}}}. This over-prediction in efficiency by Geant4 than the experimental efficiency may be due to the inaccurate dimensions of internal structure of the detector used in the simulations. The structural dimensions of the detector involves crystal radius, crystal length, dead layer thickness, Al end cap thickness and Al end cap to crystal distance ($d_{alc}$). As detector dimensions provided by the manufacturer may not be accurate or it might be possible to change with the period of time, one needs to optimize these parameters using the measured efficiency data for all the distances. It is clear that the crystal radius has uniform effect on each $\gamma$-ray photopeak where as the crystal length affects the photopeaks due to high energy $\gamma$-rays {\color{blue}\textbf{\cite{Chhavi2011}}}. It is discussed in {\color{blue}\textbf{Ref. \cite{Huy2007}}} that a long period of operation may increase the dead layer thickness of the detector multiple times than of its original thickness. Optimizing the dead layer thickness of the detector is important as it is found to largely affect the low energy $\gamma$-ray photopeaks. Proper knowledge of Al end cap to crystal distance (d$_{alc}$) is necessary to reduce the uncertainty in the source to detector distance (d). A small variation in this parameter may cause a significant variation in efficiencies in close geometry measurements. The detector parameters were tuned systematically and optimized within the limits of uncertainty using measured efficiencies. 

\begin{figure*}
    \centering
    \includegraphics[scale=0.6]{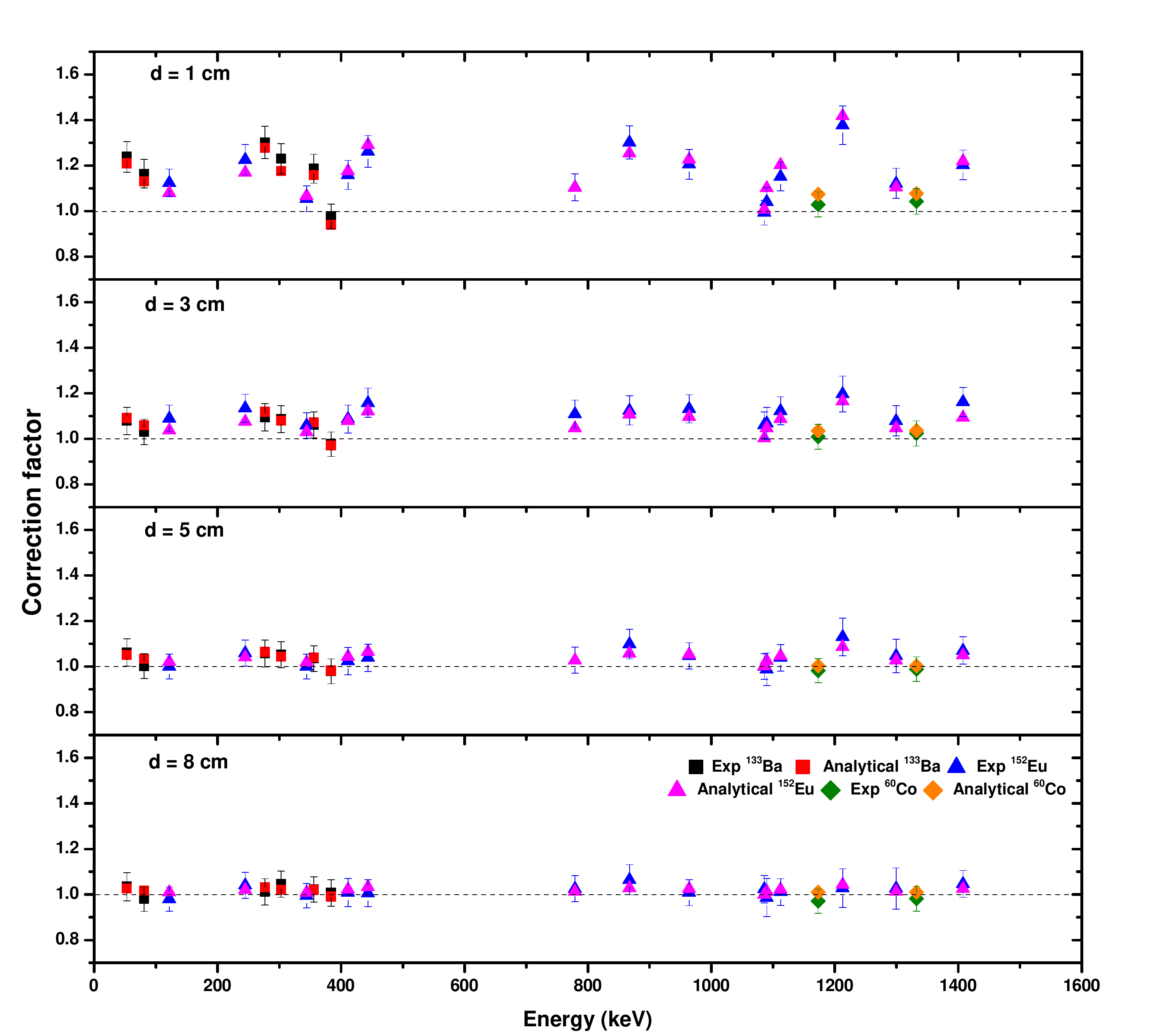}
    \caption{Experimental and analytical correction factors at d = 1, 3, 5, 8 cm}
    \label{fig:correctionfactorplot1-8}
\end{figure*}

The efficiencies calculated from the Geant4 simulations using the optimized parameters are shown in  {\color{blue}\textbf{Fig. \ref{fig:monoenergetic}}}, which well reproduces the experimental efficiencies at each source to detector distances within the error bars. The same detector parameters were used to simulate the $\gamma$-ray photopeak efficiency corresponding to multi-energetic source $^{60}$Co, $^{133}$Ba and $^{152}$Eu also. Since the $\gamma$-ray efficiencies predicted by Geant4 are independent of decay scheme of the source, it is free from the coincidence summing effect. Hence the ratio of the efficiency obtained form the Geant4 simulation to that from the measurement is the experimental correction factor ($k^{Exp}_{TCS}$) for coincidence summing effect, which is given by,

\begin{equation}
    k^{Exp}_{TCS}=\frac{Geant4~efficiency}{Experimental~efficiency}
\end{equation}

The experimental efficiency and Geant4 efficiency for $\gamma$-ray photopeak energy at each source to detector distances are shown in {\color{blue}\textbf{Figs. \ref{fig:ExpnG4}(a-f)}}. From this figure, it is clearly seen that the Geant4 simulated efficiency using the optimized detector parameters for all $\gamma$-ray photopeak energy is in good agreement with the measured efficiencies at large source to detector distances for d = 8 to 25 cm. However, a significant deviation of Geant4 efficiency can be observed from the measured values when the source to detector distance approaches smaller values from d = 8 cm and the magnitude of deviation is found to be larger at d = 1 cm, which again confirm the coincidence summing at close geometry measurements. It may be noticed from the {\color{blue}\textbf{Figs. \ref{fig:ExpnG4}(a-f)}} that Geant4 simulated efficiency match well to the experimental efficiency obtained using monoenergetic sources at all the measured source to detector distances.

The experimental values of TCS correction factor for the $\gamma$-ray photopeaks of $^{60}$Co, $^{133}$Ba and $^{152}$Eu radioactive sources are given in {\color{blue}\textbf{Table \ref{tab:correctionfactor}}}. The coincidence correction factor for these radioactive sources were also calculated using analytical method at each source to detector distance  as discussed in {\color{blue}\textbf{Section \ref{sec:Introduction}}}. The probability ($p_i$) of coincidence of two cascading $\gamma$-rays was calculated using appropriate parameters obtained from the decay schemes of the radioactive sources available in the literature{\color{blue}\textbf{\cite{Lederer1978}}}. The internal conversion coefficients, fluorescence yield and k-capture probability were taken from ref. {\color{blue}\textbf{\cite{recomendeddata}}}. The photopeak and total efficiency for each $\gamma$-ray of interest and coincident $\gamma$-ray were computed within the framework of Geant4 simulation. Coincidence summing corrected experimental efficiency is shown in {\color{blue}\textbf{Figs. \ref{fig:AnanG4}(a-f)}} for each distance as function of $\gamma$-ray energy along with the Geant4 simulated values. The correction factors obtained from the analytical method for $\gamma$-ray photopeak of $^{60}$Co, $^{133}$Ba and $^{152}$Eu radioactive sources have been given in   {\color{blue}\textbf{Table \ref{tab:correctionfactor}}}.

The correction factors estimated using the experimental and analytical method are shown in  {\color{blue}\textbf{Fig. \ref{fig:correctionfactorplot1-8}}} for all the source to detector distances chosen in this work. It is observed that the correction factors obtained from both the methods lies very close to unity for d = 8 cm. As the source to detector distance decreases further, we notice that the deviation of correction factor from unity is significant and it is predominant at d = 1. It is evident from {\color{blue}\textbf{Fig. \ref{fig:correctionfactorplot1-8}}} that the correction factor increases with decrease in the source to detector distance. It is also found that, the BEGe detector may be suitable to use in $\gamma$-ray spectroscopy experiments without considering the effect of coincidence summing if the detector is at 8 cm or above from the gamma emitting source. 

\section{Conclusion}
\label{sec:conclusion}
The FEP efficiency measurement of an electrically cooled BEGe detector has been carried out using standard mono-energetic and multi-energetic point sources as well as fabricated mono-energetic sources for different source to detector distances. The detector response was simulated and FEP efficiency was also obtained using Geant4 Monte Carlo code. The internal structural dimensions of the detector were optimized using measured efficiencies obtained from the mono-energetic sources as the detector specification provided by the manufacturers are insufficient. It is found that for the source to detector distance from 8 cm and above, there were no effect of coincidence summing in the spectra. However as the source to detector distance decreases below 8 cm, the presence of coincidence summing is evident. The true coincidence summing correction factors for radioactive sources $^{60}$Co, $^{133}$Ba and $^{152}$Eu have been determined using experimental and analytical method for BEGe detector at different source to detector distances. The coincidence summing is a very important effect and must be taken into account when performing close geometry measurements for source emitting gamma rays in cascade. The highest correction factor was observed for 1212 keV $\gamma$-ray photopeak energy of $^{152}$Eu radioactive source and it was estimated to be 1.418 using analytical method and 1.377 using experimental method. No summing correction is required for the BEGe detector if the detector is at 8 cm or above from the $\gamma$ emitting sources.

\section{Acknowledgment}
\label{sec:Acknowlegment}
The authors gratefully acknowledge the cyclotron crew of VECC, Kolkata for providing excellent quality beam throughout the irradiation experiment. One of the authors (A.G.) acknowledge the useful discussions with Prof. Supratik Mukhopadhyay and Subhendu Das regarding simulations.

\printcredits

\bibliographystyle{elsarticle-num}

\bibliography{cas-refs}


\end{document}